\documentclass[aps,prb,reprint,twocolumn,amsmath,amssymb,superscriptaddress,eqsecnum,floatfix,scrartcl]{revtex4-1}
\pdfoutput=1      
\usepackage{amssymb}  
\usepackage{color,soul} 
\usepackage{graphicx}  
\usepackage[tight]{subfigure}  
\usepackage{mathrsfs}
\usepackage[pdftex,pdfusetitle,bookmarks=true,colorlinks=true,citecolor=blue,urlcolor=blue,linkcolor=magenta]{hyperref}
\usepackage{hypcap}
\usepackage{braket}
\usepackage{xfrac}
\usepackage{tikz}
\usepackage{url}
\usepackage{setspace}
\usepackage{enumerate}
\usepackage{bm}
\usepackage{epstopdf}
\usepackage{amsmath}
\usepackage{amsfonts}
\usepackage{mhchem} 

\newcommand{\ie}{\textit{i.e.}}

\newcommand{\up}{\mathord{\uparrow}}
\newcommand{\dn}{\mathord{\downarrow}}

\renewcommand{\vec}[1]{\boldsymbol{\mathbf{#1}}}

\graphicspath{{./}{./images/}}

\begin{document}
\title{Quantum chaos dynamics in long-range power law interaction systems}

\author{Xiao Chen}
\affiliation{Kavli Institute for Theoretical Physics, University of California, Santa Barbara, CA 93106, USA}
\author{Tianci Zhou}
\affiliation{University of Illinois, Department of Physics, 1110 W. Green St. Urbana, IL 61801 USA}
\affiliation{Kavli Institute for Theoretical Physics, University of California, Santa Barbara, CA 93106, USA}

\date{\today}
\begin{abstract}
We use out-of-time-order commutator (OTOC) to diagnose the propagation of chaos in one dimensional long-range power law interaction system. We map the evolution of OTOC to a classical stochastic dynamics problem and use a Brownian quantum circuit to exactly derive the master equation. We vary two parameters:  the number of qubits $N$ on each site (the onsite Hilbert space dimension) and the power law exponent $\alpha$. Three light cone structures of OTOC appear at $N = 1$: (1) logarithmic when $0.5<\alpha\lesssim 0.8$,  (2) sublinear power law when $0.8 \lesssim  \alpha \lesssim 1.5$ and (3) linear when $\alpha \gtrsim 1.5$. The OTOC scales as $\exp(\lambda t)/x^{2\alpha} $ and $t^{2 \alpha / \zeta} / x^{ 2 \alpha} $ respectively beyond the light cones in the first two cases. When $\alpha \geq 2$, the OTOC has essentially the same diffusive broadening as systems with short-range interactions, suggesting a complete recovery of locality. In the large $N$ limit, it is always a logarithmic light cone asymptotically, although a linear light cone can appear before the transition time for $ \alpha \gtrsim 1.5$. This implies the locality is never fully recovered for finite $\alpha$. Our result provides a unified physical picture for the chaos dynamics in long-range power law interaction system. 
\end{abstract}

\maketitle

\section{Introduction}

Quantum many-body chaos has been a subject of continuous interest in the past years and has drawn a lot of attention from various subfields of physics. The dynamics of chaos can by diagnosed by the out-of-time-order commutator (OTOC)\cite{larkin1969}, 
\begin{align}
\label{eq:Ct}
C(t)=-\langle [\hat O_1(t), \hat O_2]^2\rangle_{\beta}
\end{align}
which measures the non-commutativity between a Heisenberg operator $\hat O_1(t)=e^{i\hat Ht}\hat O_1 e^{-i\hat Ht}$ and a time independent simple operator $\hat O_2$. This quantity has a natural classical origin, with the commutator becoming a Poisson bracket, which measures the separation of nearby trajectories in the flow of the dynamical system. In the classical chaotic system, the separation grows exponentially in time and the growth rate is called the Lyapunov exponent. This sensitivity to initial condition is commonly known as the butterfly effect. 

In quantum system, the unitarity and quantum effect can produce different scaling behaviors in OTOC. 

For instance, in some many-body chaotic system with all-to-all interactions, $\hat O_1(t)$ spreads out extremely fast in Hilbert space and OTOC can grow exponentially in time, \ie, $C(t)\sim \exp(\lambda t)$. $\lambda$ here is a quantum analogy of the Lyapunov exponent and  characterizes the quantum butterfly effect at early time\cite{hayden2007,sekino2008, kitaev2015,roberts2015b,maldacena2016}. These systems in the literature are referred to as ``fast scramblers"\cite{sekino2008}.

Nevertheless, typical many-body quantum system do not have all-to-all interactions. The spatial locality puts extra constraints on the quantum dynamics. Consider a Heisenberg operator $\hat O_1(t)$ initially supported only at origin. As time evolves, the size of the operator grows. This can be measured by its OTOC with another operator $\hat O_2(x)$ sitting at spatial coordinate $x$. In systems with local interactions, $C(x,t)$ is zero at $t=0$ and starts to become appreciable at time  $t= x/v_B$. Here $v_B$ is the so-called butterfly velocity\cite{shenker2013a,Roberts2015}, which characterizes the ballistic spreading of chaos. 

In the last several years, the specific scaling form of $C(x,t)$ has been extensively investigated across a wide variety of systems with local interaction. In systems with large $N$ limit, field theory calculations indicate that $C(x,t) $ forms a ballistic ``front'' which is approximately $\exp(\lambda(t-x/v_B))$ when $x\gtrsim v_Bt$\cite{shenker2013a, Roberts2015, gu2017}. The large $N$ limit allows enough room for the Heisenberg operator $\hat O_1(t)$ to grow in local Hilbert space and leads to an exponential growth of $C(x,t)$ for an extended period of time. In comparison, it does not appear to have such Lyapunov regime for systems with small onsite Hilbert space dimension, at least at infinite temperature. Ref.\ \onlinecite{keyserlingk2017,nahum2017} design a one dimensional local Haar random circuit model and analytically show that  $C(x,t) \sim \text{erfc}(a(x-v_Bt)/\sqrt{t})$ when   $x \sim v_B t$ . This diffusive wavefront, meaning the width broadens as $\sqrt{t}$, is further confirmed in the numerics of realistic quantum spin-$1/2$ chain models\cite{leviatan2017,xu2018}. Here $v_B$ is no greater than the Lieb-Robinson velocity that appears in the Lieb-Robinson bound\cite{lieb1972}, which points out the information can at most spread linearly in systems with local interactions.

The physics could be  different in systems whose interaction decays as $r^{ - \alpha}$ as a function of interaction distance $r$. The range of this type of interaction interpolates between all-to-all and local interactions we mentioned before. They exist in a wide variety of experimental platforms, such as ultracold atoms\cite{Blatt_2012}, trapped ions\cite{Bloch_2012} and solid state spin defects\cite{Awschalom_2012}. The information propagation in these systems could potentially be much faster than systems with short-range interaction. In the past decade, a tremendous amount of effort has been devoted to derive a tight Lieb-Robinson bound\cite{lieb1972} to power law interaction systems. Hastings and Koma first generalized the method in Ref.~\onlinecite{lieb1972} and proved a logarithmic light cone bound for $\alpha>D$, where $D$ is the spatial dimension\cite{Hastings2006}. However, this bound is not tight for large $\alpha$. A series of subsequent improvements were proposed and proved in Ref.~\onlinecite{Gong2014,Foss-Feig2015,Storch_2015,matsuta_improving_2017,Tran2018,else_improved_2018}. As of now\cite{Tran2018,else_improved_2018}, we have a power law light cone bound for $\alpha>2D$ and this light cone asymptotically becomes linear when $\alpha\to\infty$. So when $\alpha > 2D$, a local perturbation can spread out at most algebraically rather than exponentially in time.

In this paper, we deal with the problem of quantum chaos in power law interaction system. We aim to obtain the light cone structure of chaos dynamics and scaling forms of $C(x, t)$, for which the generic bound above can not answer precisely. The crucial difficulty in dealing with this problem comes from two aspects: First, chaotic system is non-integrable and analytical treatment is intrinsically hard. Meanwhile, numerical calculation based on either exact diagonalization or matrix product approach\cite{Verstraete_2008} usually limits to small system size due to the large entanglement generated by chaotic dynamics. Additionally, long range interaction generates stronger finite size effect\cite{chen_measuring_2017, Luitz2018} compared to the models with short range interaction. To circumvent these difficulties, we construct a Brownian quantum circuit model with power law interaction. It keeps the power law decay strength of the interaction, while dispenses its particular form by replacing it with a noisy evolution. This is in spirit similar to the local Haar random circuit\cite{keyserlingk2017,nahum2017} which successfully describes the hydrodynamics of the chaos propagation in a generic {\it locally} interacting chaotic systems. Hence we expect our Brownian circuit model serves to be a minimal model for the long-range interacting chaotic system, which uncovers universal features of various quantities in chaos dynamics. Here the random nature of the interaction allows us to express the ``height distribution" of the operator evolution in terms of a closed-form master equation\cite{lashkari2013,zhou_operator_2018,xu_locality_2018}. We therefore map the complex quantum dynamics to a relatively simple classical stochastic problem. Although complete analytical solution is still not available except for $\alpha=0,\infty$, the master equation and the associated stochastic process are intuitively simple and allow analytical arguments and large scale numerical simulations on {\it thousands} of sites to greatly reduces the finite size effect. 


The Brownian circuit model we study is in one dimension with each site hosting $N$ qubits. Our focus is on small $N$ and large $N$ limits. When $N=1$, we will demonstrate the emergence of the {\it linear}, {\it power law} and {\it logarithmic} light cones from the perspective of chaos propagation. For the linear light cone regime, we can define a constant butterfly velocity $v_B$ to characterize the speed of chaos propagation. For the power law and logarithmic light cone regime, we generalize the butterfly velocity $v_B$ to be a time dependent quantity $v_B(t)$, which is the derivative of the light cone trajectory $x_{\rm LC}(t)$. We show that when $0.8\lesssim\alpha\lesssim1.5$, $C(x, t)$ has a power law light cone, in which $v_B(t)$  grows algebraically in time. $C(x, t)$ beyond the light cone is a power law function in both spatial and temporal direction, i.e., $C(x,t)\sim t^{\frac{2\alpha}{\zeta}}/x^{2\alpha}$. When $0.5< \alpha \lesssim0.8$, we enter into the logarithmic light cone regime, in which $v_B(t)$ grows exponentially in time. When $\alpha<0.5$, the model completely loses locality and the physics is essentially the same as $\alpha=0$ limit, \ie, all to all interactions.


The results above do not violate the current information bound for power law interaction systems\cite{Tran2018}. Instead, it suggests room to potentially tighten the power law bound for $\alpha > 2D$ in Ref.~\onlinecite{Tran2018}. For example, we observe the emergence of the {\it linear} light cone when $\alpha\gtrsim 1.5$, meaning a time independent $v_B$ even in long-range interaction system. In addition, we find that when $\alpha\geq 2$, the OTOC has a diffusive wave front, the same as systems with local interaction\cite{keyserlingk2017,nahum2017}. 

\begin{figure}[hbt]
\centering
 { \includegraphics[width=.48\textwidth]{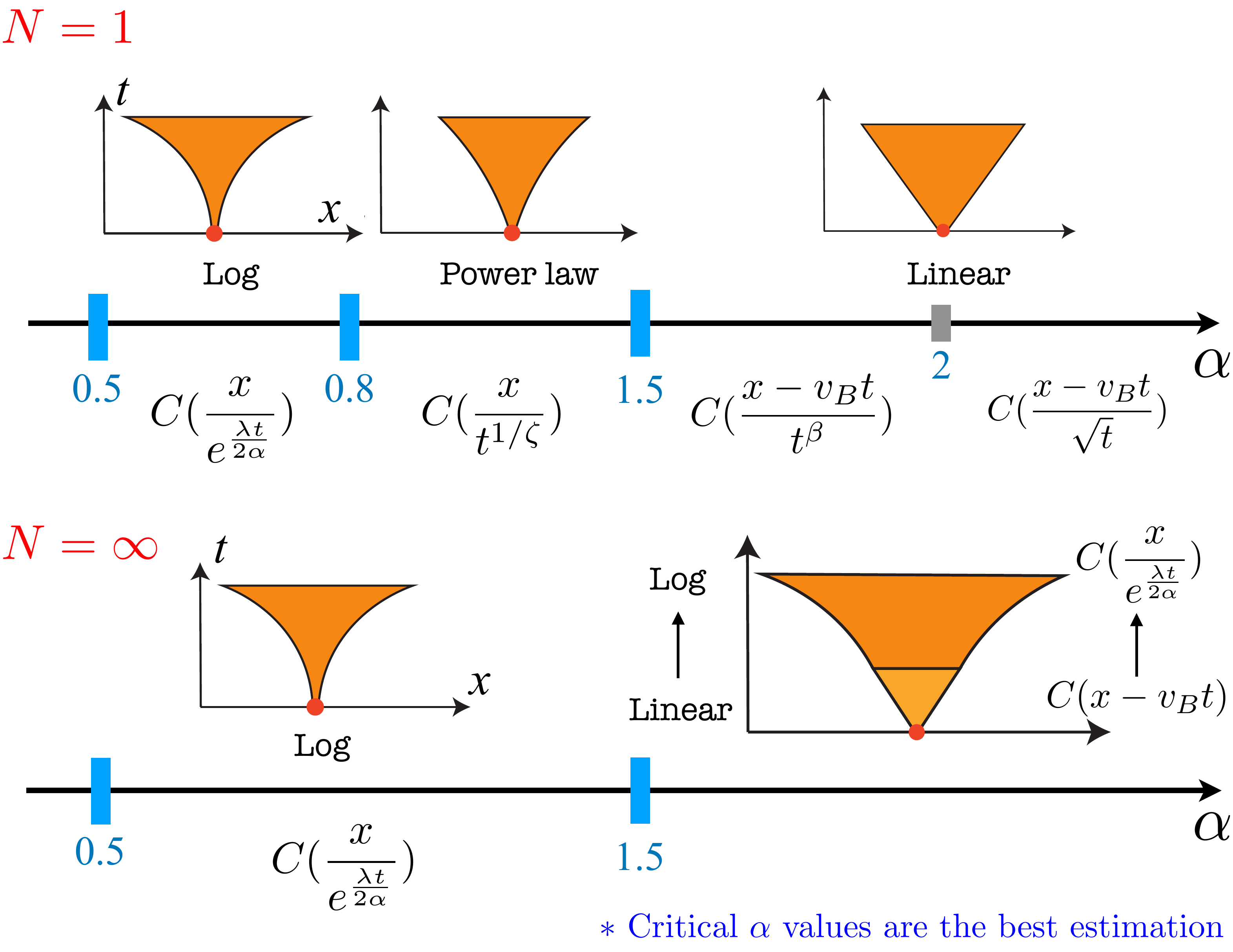}}
\caption{ Upper panel: The light cone structures and the scaling forms of OTOC when varying $\alpha$ at $N=1$. In the power law light cone regime between 0.8 and 1.5, the OTOC grows as algebraically in time. The linear light cone regime with $\alpha\gtrsim 1.5$ can be further divided into two subregimes at $\alpha = 2$ according to the different scaling forms of OTOC. Lower panel: The light cone structure and scaling form of OTOC when varying $\alpha$ in the large $N$ limit. The values of the critical $\alpha$ are estimated from the numerics.}
\label{fig:schematic}
\end{figure}

Besides $\alpha$, the light cone structure also depends on the parameter $N$, which controls the dimension of the onsite Hilbert space. In the large $N$ limit, we show that the OTOC is described by a fractional Fisher-Kolmogorov-Petrovsky-Piskunov (FKPP) equation, which exhibits a logarithmic light cone structure in the regime $0.5 \lesssim \alpha \lesssim 1.5$. When $\alpha \gtrsim 1.5$, we observe  a two-segment light cone from linear to logarithmic as time evolves. We further discuss the scaling forms of OTOC in different light cone regimes and summarize the main results in Fig.~\ref{fig:schematic}.


The rest of the paper is organized as follows. In Sec.~\ref{sec:master_equ}, we discuss Brownian quantum circuit and derive the master equation governing the operator growth in systems with power law interactions. In Sec.~\ref{sec:N_1}, we perform numerical simulation on the $N=1$ case and demonstrate the appearance of various light cone structure as we tune $\alpha$. We then perform data collapse on $C(x,t)$ and discuss possible scaling functions of OTOC in Sec.~\ref{sec:N_1_OTOC}. In Sec.~\ref{sec:N_infty}, we discuss the light cone structures and scaling functions of OTOC in the large $N$ limit and compare the results with the small $N$ limit. Finally, we summarize our results and discuss possible future directions in Sec.~\ref{sec:conclusion}.

\section{The Brownian quantum circuit and master equation}
\label{sec:master_equ}
We begin by introducing the dynamics in the operator space. Let $B_\mu $ be a complete orthonormal operator basis, any operator can be expanded as 
\begin{align}
O(t)=\sum_\mu \alpha_\mu(t) B_\mu. 
\end{align}
Under {\it unitary} time evolution,  $|\alpha_\mu(t)|^2$  can be interpreted as the probability of the basis $B_\mu$, where the total probability for properly normalized operator $\sum_\mu  |\alpha_\mu(t)|^2=1$ is conserved. 

In a generic chaotic system, the operator dynamics is complicated and it is usually hard to keep track of the evolution of each component $|\alpha_\mu (t)|^2$. It is also unnecessary to know each component of $|\alpha_\mu (t)|^2$ if we only focus on the universal information in OTOC. For instance the insensitivity to the choice of local operator $V$ suggests that we can combine $|\alpha_\mu (t)|^2$ on each site and study the coarse grained hydrodynamics. Inspired by the local random Haar circuit models\cite{keyserlingk2017,nahum2017}, we introduce a model that is maximally symmetric on different basis after random averaging to simplify the dynamics. 


The model we consider has $L$ sites, each of which is a quantum dot that hosts $N$ spin-$\frac{1}{2}$ degrees of freedom. The model contains only the couplings between spins from different dots. To be concrete, let $\sigma^{\mu}( i, n ) $ be the Pauli sigma matrix of $n$th spin at dot $i$, and the Hamiltonian is composed of two-body interactions, 
\begin{equation}
\label{eq:H_brownian}
H = \sum_{i\ne j} \sum_{n,m=1}^N  \sum_{\mu, \nu = 0}^3   J_{\mu \nu}(i, j, m, n,  t) \sigma^{\mu}( i, m )  \sigma^{\nu}( j, n ) .
\end{equation}
where the coupling constant $J$ is proportional to a power function of the distance $\frac{1}{| i - j|^{\alpha}}$. In order to make the model tractable, we take the couplings to be independent Brownian motions. Dividing the evolution into short periods of $\Delta t$ intervals, the coupling at the $s$-th interval is approximately
\begin{equation}
J_{\mu \nu } ( i, j, m, n, t = s \Delta t   ) = \Delta B^s_{i,j,m,n} (t ) \frac{g}{| i - j|^{\alpha} }, \,\, g = \sqrt{ \frac{1}{8}} ,
\end{equation}
where $\Delta B^s_{i,j,m,n}( t )$ are independent Gaussian random numbers with variance proportional to $\Delta t $. The complete time evolution is generated in the continuum limit of
\begin{equation}
e^{ -i H_s \Delta t } e^{ -i H_{s-1}  \Delta t }  \cdots .
\end{equation}
This type of model is called the Brownian quantum circuit. The time evolution is a random walk on the unitary group in the direction of allowed couplings. The statistical average of the operator spreading is analytically tractable and many of the variants have been used to study the quantum dynamics in chaotic systems\cite{lashkari2013,xu2018,zhou_operator_2018, Gharibyan2018}.

In the one dimensional model we consider, the operator dynamics is fully determined by the operator height distribution function
\begin{align}
f({\vec h},t)=\sum_{{\rm height}(B_\mu)=\vec{h}}|\alpha_\mu(t)|^2, 
\end{align}
where height ${\vec h}=(h_1,h_2,\cdots,h_L)$ is a $L$-component vector. The height of a Pauli basis $B_{\mu}$ on each site is the number of non-identity operators therein.  The distribution function $f({\bf h})$ groups all the probability contributions whose corresponding basis has the same height $\vec{h}$. The height distribution is important since it contains all the necessary information of operator growth and the mean height is equal to OTOC\cite{roberts2018, xu2018, zhou_operator_2018}. In our previous paper \cite{zhou_operator_2018}, we studied a single Brownian quantum dot of $N$ qubits with all-to-all interaction, and we derived the master equation of $f(h)$ with $h\in [0, N]$ as the height distribution on a single dot. Following a similar method, we can show that the evolution of the joint distribution $f(\bf h)$ in a one dimensional model is governed by the master equation 
\begin{equation}
\label{eq:master-N}
\begin{aligned}
\frac{\partial f( \vec{h}, t ) }{\partial t}  =&   \sum_{j\ne i } 3 D_{ij} h_{j} (N-h_i+1)f( \vec{h} - \vec{e}_i , t ) \\
&+ \sum_{j\ne i}  D_{ij} h_{j}(h_{i}+1) f( \vec{h} + \vec{e}_i , t ) \\
 & - \left\{  \sum_{j\ne i } 3 D_{ij} h_{j} (N-h_i) +  D_{ij} h_ih_{j} \right\} f( \vec{h}, t )  ,
\end{aligned}
\end{equation}
where the coefficient $D_{ij} $ is $ \frac{1}{|i - j|^{ 2\alpha} }$. ${\bf e}_i$ is a L-component vector which takes unit value at site $i$ and is equal to zero at other sites.  If we take $D_{ij}$ to be short-range interaction, the equation is similar to the one derived in Ref.~\onlinecite{xu2018}. Starting from $f({\bf h})$, we can compute the mean height $\overline{h}_j$ at each site and obtain the spatial and temporal profile of OTOC. For instance, if we take $\hat{O}_1(t = 0) $ and $\hat{O}_2$ as simple operators at dot $i$ and $j$ respectively, the mean height $\overline{h}_j$ is exactly the same as OTOC $C(x=|i-j|, t )$ defined in Eq.~\eqref{eq:Ct}.

\begin{figure}[h]
\centering
\includegraphics[width=\columnwidth]{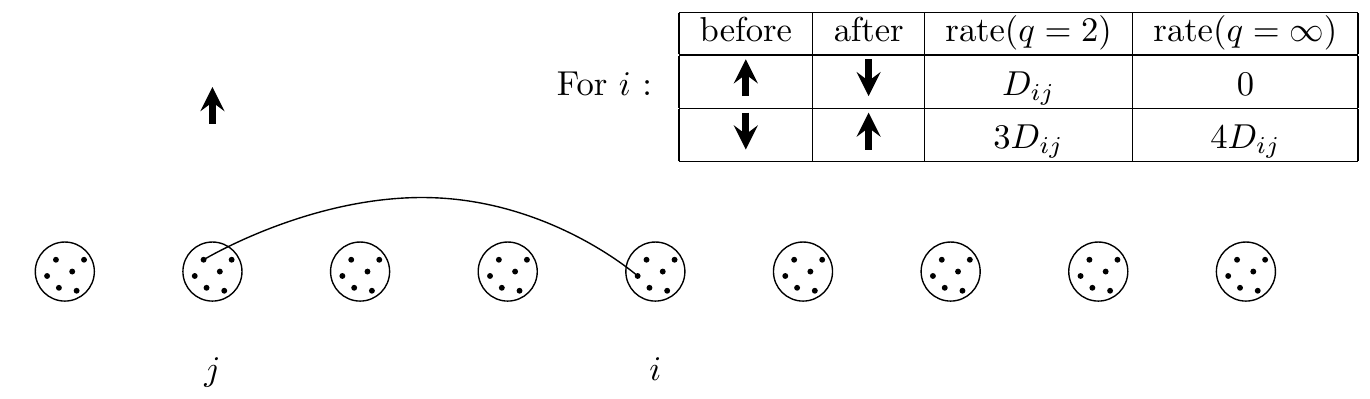}
\caption{Transition rate induced by an interaction term between spins from dot $j$ and dot $i$. The flipping rate of $N=1$ case (shown in the table) for spin at dot $i$ is only nonzero when spin $j$ is up; the rate for flipping up is higher, giving rise to the operator spreading. }
\label{fig:trans_rate}
\end{figure}

We first deal with the case of $N = 1$, representing spin chain model with small onsite Hilbert space dimensions. The height at each dot can only take $0$ or $1$ and therefore the model is equivalent to a non-equilibrium kinetic Ising model\cite{Hohenberg1977}. In this case, the master equation becomes, 
\begin{equation}
\label{eq:master-N1}
\begin{aligned}
\frac{\partial f( \vec{h}, t ) }{\partial t}  =& \sum_{j \ne i }  3 D_{ij} h_{j} f( \vec{h} - \vec{e}_i , t ) +  \sum_{ j \ne i } D_{ij} h_{j} f( \vec{h} + \vec{e}_i , t ) \\
 & - \left\{ \sum_{j\ne i } 3 D_{ij} h_{j} (1-h_i) +  D_{ij}   h_i  h_{j} \right\} f( \vec{h}, t )  .
\end{aligned}
\end{equation}
This equation describes a Markov process shown in Fig.~\ref{fig:trans_rate} where $\uparrow$ denotes $h=1$ and $\downarrow$ denotes $h=0$ configuration. The two-site interaction between $i$ and $j$ induces a transition rate of height change: the dot $j$ with height $h_j = 1$ will increase the height $h_i$ from $0$ to $1$ with rate $3 D_{ij}$ while decrease its height from $1$ to $0$ with rate $D_{ij}$. Whereas if $h_j=0$, the transition for $h_i$ to go from one configuration to another is always zero. Such kind of update rule for Ising spin dynamics is the same as the one spin facilitated Fredrickson-Andersen model, which is used to study the dynamics of classical spin glass\cite{Fredrickson1984}. Here the corresponding quantum qubit on each site has Hilbert space dimension $2$.  This result can be further generalized to a $q$ dimensional local Hilbert space and the transition rates becomes $4 (1 - \frac{1}{q^2}) D_{ij}$ and $\frac{4}{q^2}D_{ij}$. In the $q\to\infty$ limit, the rate of flipping $\uparrow$ to $\downarrow$ is zero. The update rule is simplified while the physics remains the same. After sufficiently long time evolution, the system reaches the final steady state with uniform mean height $\overline{h(x,t\to\infty)}=1-1/q^2$.

The dynamics of the master equation in Eq.\eqref{eq:master-N1} has two simple limits that have been studied in the previous literatures. 

When $\alpha = 0$, the interaction is equally weighted between any pair of quantum dots. Therefore we can view this as an effective single quantum dot with $L$ spins. Because all dots are identical, the full distribution only depends on the total height $\sum_{i} h_i $ (which is the height of the effective single dot). The OTOC has an initial exponential growth and the full dynamics can be described by a general logistic function behavior\cite{chen_operator_2018, zhou_operator_2018}. The same scaling behavior should hold even when $\alpha$ is close to $0$ that the locality is completely lost.

\begin{figure}[h]
\centering
\includegraphics[width=\columnwidth]{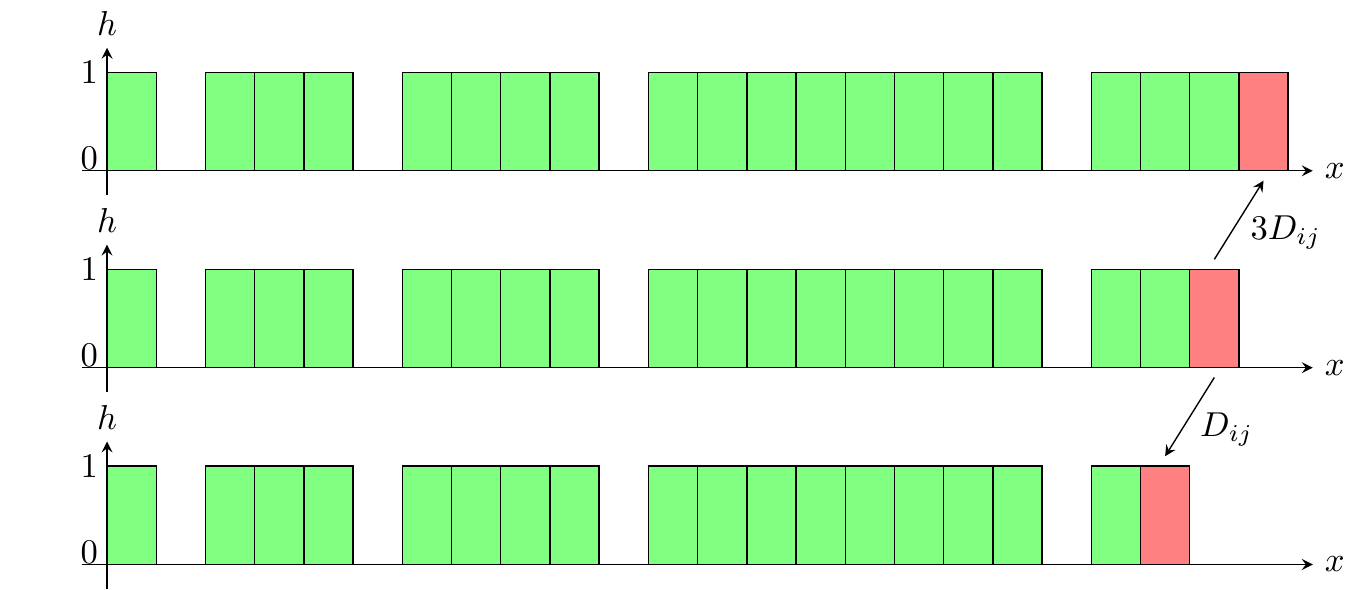}
\caption{End point of an operator. Three panels show typical height configurations at $\alpha = \infty$, where green block represents $h = 1$ and empty for $h = 0$. The red block marks the right most site with height $1$. The flip occurs at nearest neighbor site. Arrows label the transition rates from the middle state to the top and bottom ones. There are also flipping processes to the left of the end point, but they quickly equilibrate.}
\label{fig:end_point}
\end{figure}
When $\alpha = \infty$, the transition rate is restricted to the nearest neighbor dots, \ie, $D_{ij}\neq 0$ only when $i=j\pm 1$. We set the initial condition to be $h_1=1$ with rest of $h_j=0$. As shown in Fig.~\ref{fig:end_point}, a typical height configuration in this limit will have a regime with high density of $h=1$ on the left and a $h = 0$ domain on the right. The red block is the right most one with height $1$, which we will call the end point. In the $q\to\infty$ limit, one can view this as a domain wall between $h=1$ domain and $h=0$ domain. The end point performs a biased random walk\cite{nahum2017,keyserlingk2017} towards the right and the mean value $\overline{h(x,t)}$ propagates ballistically in time with the front broadening diffusively.  This wave front interpolates $h = 0$ domain and the left side of the end point, which  quickly equilibrates to have an average value of $h_{\rm sat} = \frac{3}{4}$. This gives the same picture described by the random local unitary circuit\cite{keyserlingk2017,nahum2017}. However, as we will show later, this biased random walk picture breaks down as we reduce $\alpha$ for two reasons: The end point can have non-local random walks that is not restricted to neighbors within fixed radius. The regime on the left of the end point does not immediately equilibrate after the front sweeps through. Hence the end point distribution, though can be defined, does not directly relate to the mean height or OTOC. We need to directly compute the mean height by the master equation.

The full joint distribution $f(\vec{h})$ governed by Eq.\eqref{eq:master-N1} or Eq.\eqref{eq:master-N}, resembles a many-body wave function of the tensor product of heights. Recently, Ref.~\onlinecite{xu2018} used the matrix product state (MPS) based algorithm to represent and evolve $f( \vec{h},t )$ in Brownian circuit with local interaction and studied the crossover from large $N$ to small $N$ limit. We take an alternative approach here to directly simulate the Markov process that generates and samples $f( \vec{h} )$ for $N=1$. We will use this method to analyze the resulting light cone structure, butterfly velocities and the scaling form of OTOC in Sec.~\ref{sec:N_1}.



In the end, we briefly mention the $N \rightarrow \infty$ limit. In this case, we study the normalized height $\underline{h} = \frac{h}{N}$, which can continuously vary from $0$ to $1$. The height fluctuation in the Markov process is an order $\mathcal{O}(\frac{1}{N})$ effect and will not be considered here-- we are now in the mean field limit. We can write down the evolution equation for $\underline{h}(x, t)$,
\begin{align}
\label{eq:mean_field}
  \frac{\partial \underline{h}( x, t ) }{ \partial t } =& \int dy \, \underline{h}( y, t ) D( y, x ) (1 - \underline{h}(x, t ) )\nonumber \\
  & - \frac{1}{3} \int dy \, \underline{h}( y, t ) D( y, x )  \underline{h}(x, t )\nonumber \\
=& \int dy \, \underline{h}( y, t ) D( y, x ) (1 - \frac{1}{h_{\text{sat}}} \underline{h}(x, t ) )
\end{align}
where the kernel  $D( x, y )$ is $\frac{1}{|x - y|^{2\alpha}}$. The first term represents the ``flip up'' rate generated by the portion of $\up$ spins at site $y$ to portion of $\dn$ spins at site $x$, while the second term gives the ``flip down'' rate.  This equation is a generalization of Fisher-Kolmogorov-Petrovsky-Piskunov (FKPP) equation\cite{Fisher1937, kolmogorov1937study} with diffusion kernel replaced by the power law kernel. Indeed, in the limit $\alpha\to\infty$, it reduces back to the ordinary FKPP equation with stable traveling wave solution\cite{ablowitz1979}. However, at finite $\alpha$, it can exhibit strikingly different dynamics. The full analysis of this equation will be performed in Sec.~\ref{sec:N_infty} which will give the light cone information and the spatiotemporal structure of mean height (OTOC). 


\section{Chaos dynamics at $N=1$}
\label{sec:N_1}

\subsection{The formation of light cone}
\label{sec:N_1_light_cone}

We numerically simulate the Markov process described by Eq.\eqref{eq:master-N1} with initial condition
\begin{equation}
h( x = 1, t =  0 ) = 1, \quad h( x> 1, t = 0 ) = 0 
\end{equation}
and open boundary condition. This initial condition represents a simple localized quantum operator at $t=0$. In each run of the simulation (one sample), $h(x, t)$ at fixed time $t$ is a classical configuration with $0$ or $1$ on each site. We take $L\in [10^3, 10^4]$ and average over 25000 samples to obtain $\overline{h(x,t)}$.  The large system size and sample number allow us to treat $\overline{h(x, t)}$ as a continuous function of $x$ and $t$. We take $L = 2000$ and $L = 10000$ to estimate the finite size effect. 

After sufficiently long time evolution, $\overline{h(x,t)}$ will approach the steady state value $\overline{h(x,t\to\infty)}=h_{\rm sat}=0.75$. In contrast to the common interest of Markov process, our focus here is not the steady state, but the entire relaxation dynamics towards it. In particular, we will investigate the formation of the effective light cone and the scaling form of OTOC. 

As the first step, we compute $\overline{h(x, t)}$ at $\alpha=0$ as a benchmark. In this limit, the transition probability set by $D_{ij}$ is independent of the locations of two spins and $\overline{h}$  quickly becomes uniform on each site. As we mentioned earlier, the total height is the height in the effective single quantum dot\cite{zhou_operator_2018}. Indeed the result matches if we rescale $D_{ij}$  by $L$.

\begin{figure}[hbt]
\centering
 \subfigure[]{\label{fig:alpha_05} \includegraphics[width=.48\columnwidth]{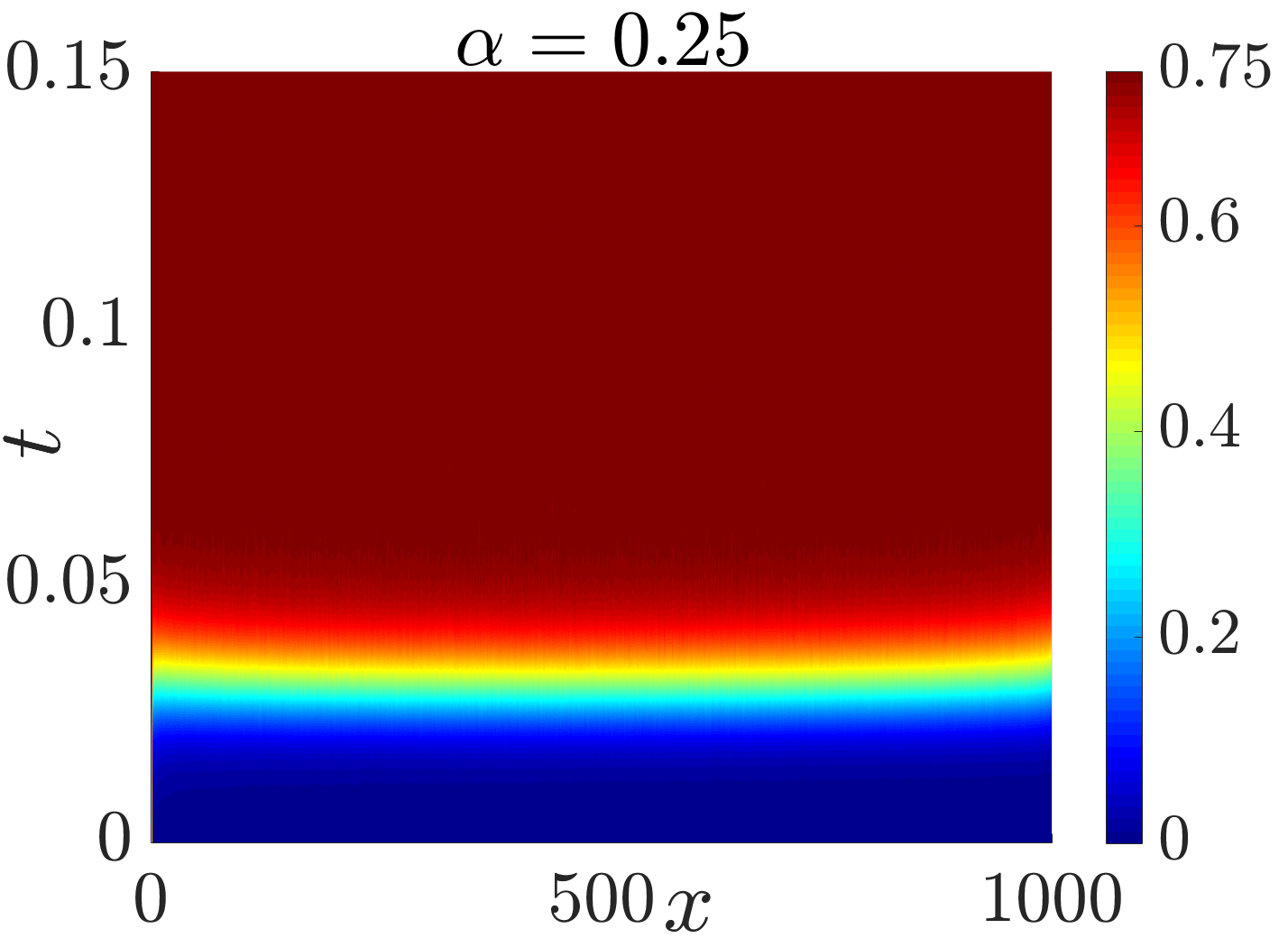}}
 \subfigure[]{\label{fig:alpha_15} \includegraphics[width=.48\columnwidth]{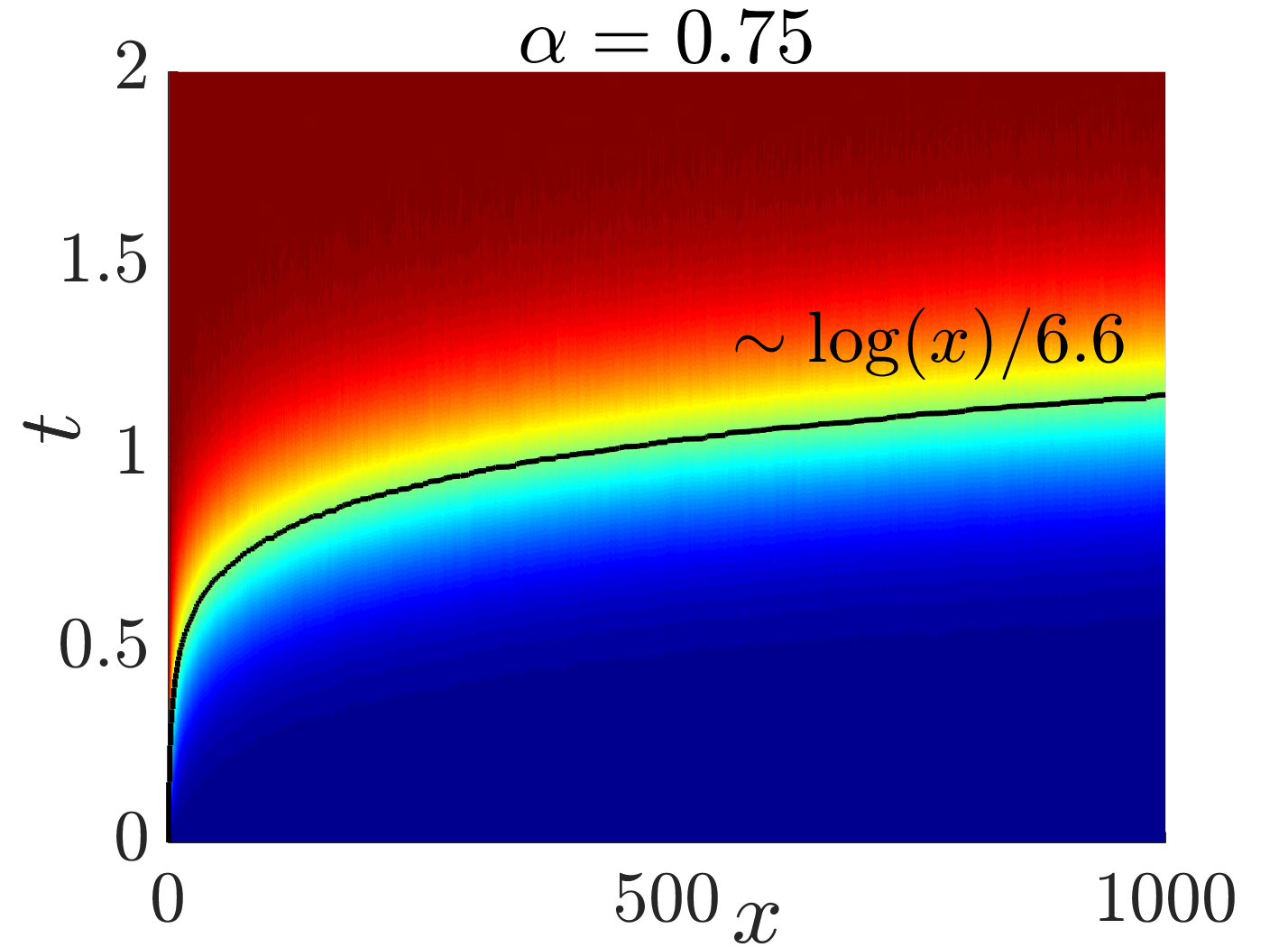}}
 \subfigure[]{\label{fig:alpha_25} \includegraphics[width=.48\columnwidth]{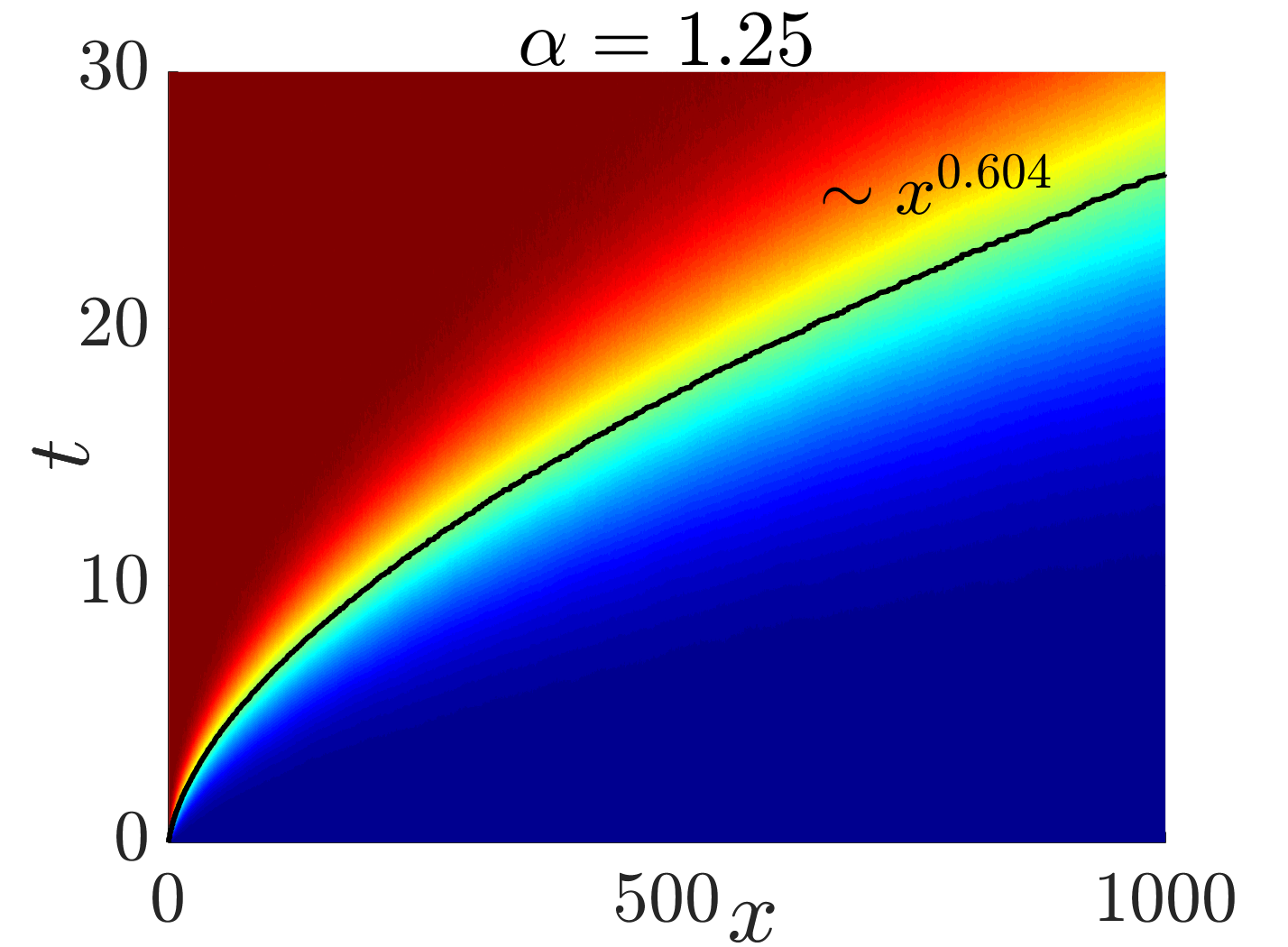}}
 \subfigure[]{\label{fig:alpha_5} \includegraphics[width=.48\columnwidth]{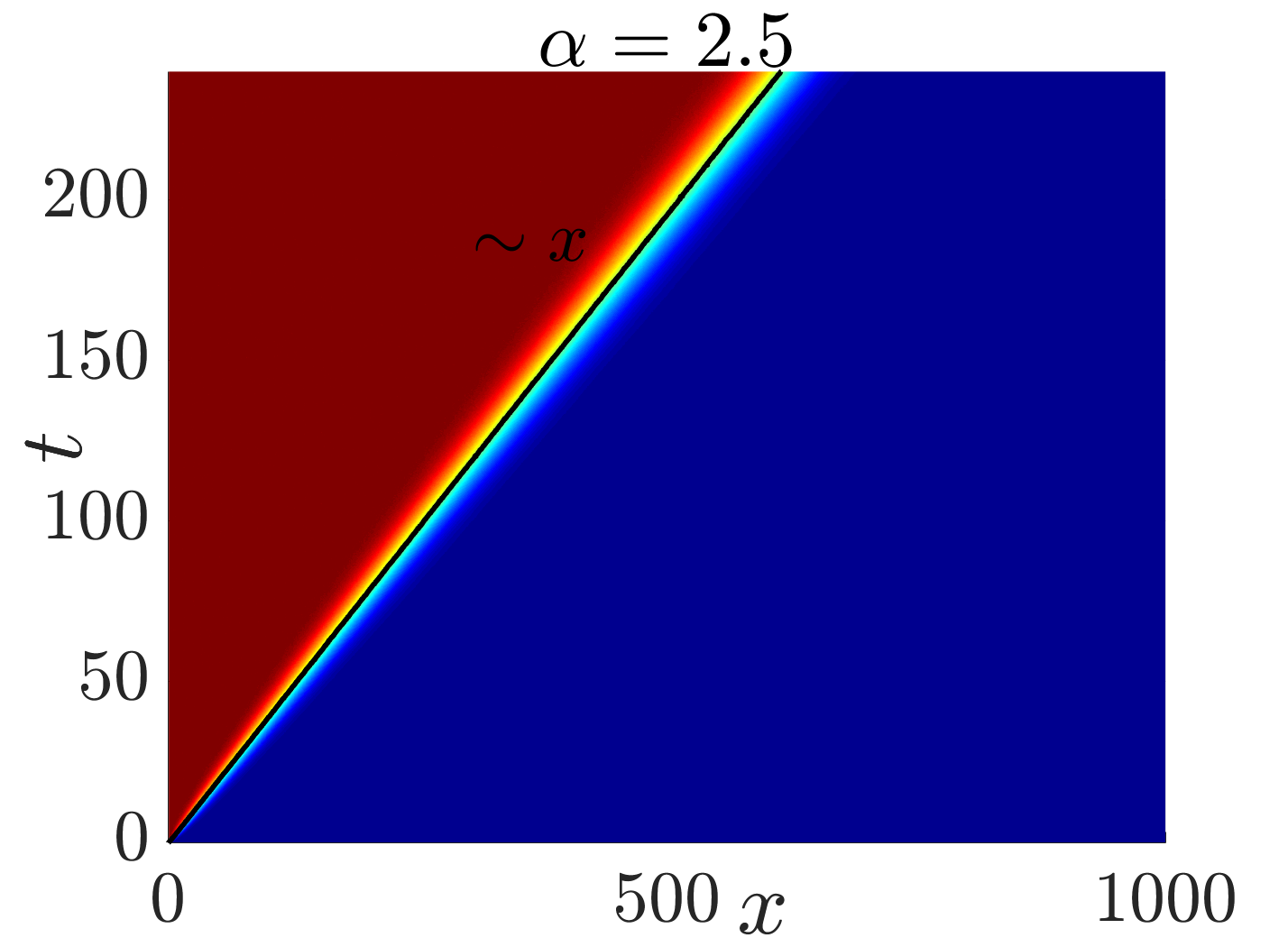}}
\caption{ The spreading of OTOC at various exponent $\alpha$. OTOC $ = h_{\rm sat}/2$ on the black line (light cone $t_{\rm LC}(x)$) in (b) (c) and (d).} 
\label{fig:alpha_contour}
\end{figure}

We now turn to the numerical simulation for $\alpha>0$. The physics of $\alpha < 0.5$ is similar to the case of $\alpha = 0$. The quantum information spreads out almost instantaneously to the entire system and therefore $\overline{h(x,t)}$ is independent of $x$. This can be clearly observed in Fig.~\ref{fig:alpha_05}, where the height simultaneously reaches $h_{\rm sat}$ everywhere in space. $\overline{h(t)}$ has an exponential growth in early time with decreasing Lyapunov exponent $\lambda$ for increasing $\alpha$.

\begin{figure}[hbt]
\centering
 \subfigure[]{\label{fig:log_cone} \includegraphics[width=.477\columnwidth]{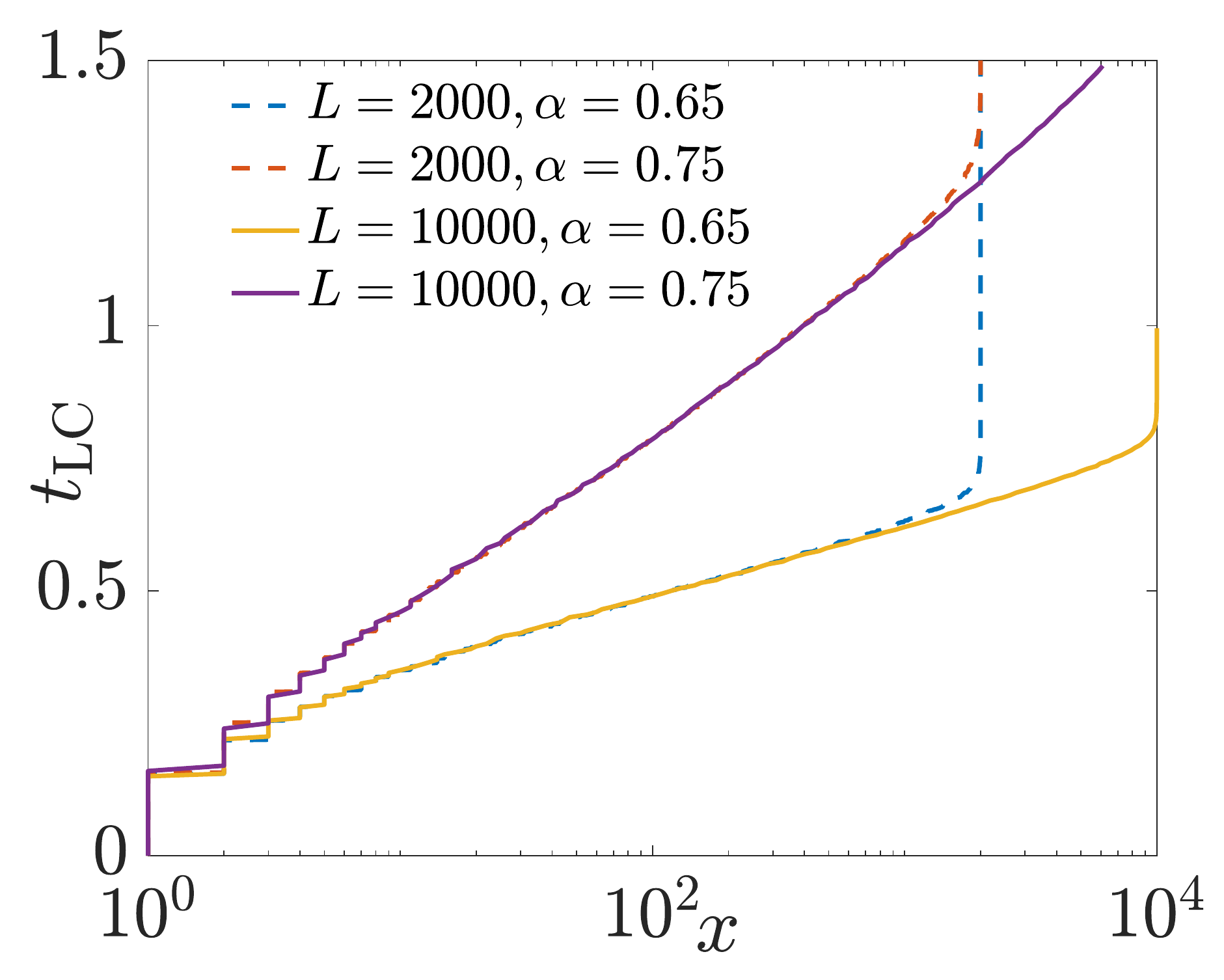}}
 \subfigure[]{\label{fig:power_cone} \includegraphics[width=.483\columnwidth]{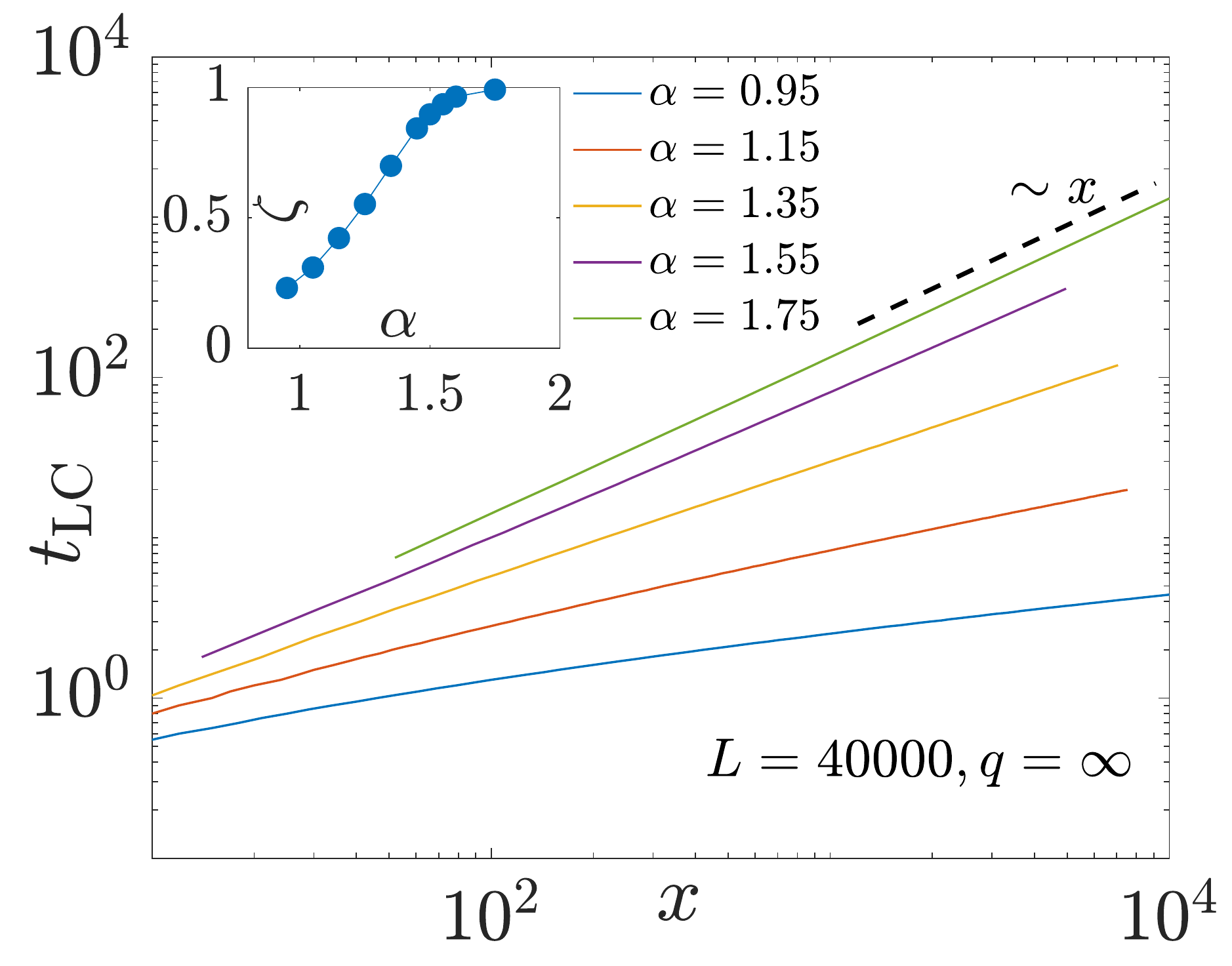}}
\caption{ Light cone scalings. (a) Linear behavior of $t_{\rm LC}$ on semi-log scale indicates a logarithmic light cone when $\alpha \lesssim 0.8$. The solid line ($L = 10000$) stretches linearly longer than the dashed line ($L = 2000$), showing that logarithmic light cone will persist in the infinite system. (b) The power law light cone $t=x^{\zeta}$ on the log-log scale for $\alpha \gtrsim 0.8$. We extract the exponent $\zeta$ at various $\alpha$ in the inset.} 
\label{fig:light_cone}
\end{figure}

The locality emerges when $\alpha>0.5$. We define the light cone to be the boundary $t_{\rm LC}(x)$, below which $\overline{h(x,t)}$ is smaller than the threshold value. In our simulation, we set this threshold to be $h_{\rm sat}/2$. We define its inverse function to be $x_{\rm LC}( t )$, thus
\begin{equation} 
  \overline{h( x_{\rm LC}(t), t )} = \overline{h( x, t_{\rm LC}(x) )} = \frac{1}{2} h_{\rm sat}.
\end{equation}
Our convention is that a logarithmic light cone corresponds to a logarithmic function $t_{\rm LC}(x)$ rather than $x_{\rm LC}(t)$.
As shown in Fig.~\ref{fig:alpha_15}, when $\alpha=0.75$, we observe that the boundary curve is a logarithmic function of $x$, indicating that the butterfly velocity $v_B(t) = d x_{\rm LC}/dt$ grows exponentially in time. The coefficient of the logarithmic curve is increasing for increasing $\alpha$ (Fig.~\ref{fig:log_cone}). 

When $\alpha \gtrsim 0.8$, we find the logarithmic light cone to be replaced by a sublinear power law light cone $t_{\rm LC} (x) \sim x^{\zeta}$ with $\zeta<1$(Fig.~\ref{fig:alpha_25}). As shown in Fig.~\ref{fig:power_cone}, we notice that the exponent $\zeta$ increases as we increase $\alpha$. When $\alpha$ exceeds about $1.5$, it becomes a linear light cone with $\zeta=1$ at large time. This suggests that the  wave front is propagating asymptotically at constant velocity $v_B$, the same as systems with local interaction. We expect that these different light cone regimes should be observed in a realistic spin chain model with power law interaction. Notice that the power law or linear light cone in the range $\alpha>1$ is far below the current information bound proposed in Ref.~\onlinecite{Tran2018}.

\subsection{The scaling form of OTOC}
\label{sec:N_1_OTOC}

To better understand the possible scaling forms of OTOC in different light cone regimes, we perform data collapse for the front of $\overline{h(x)}$ at different times for various $\alpha$. The front of the $\overline{h(x)}$ profile is the regime which interpolates the regimes of $h  = h_{\rm sat}$ and $h = 0$. We consider the following scaling ansatz $\overline{h(x,t)}$ for data collapse:
\begin{equation}
h( x, t ) \rightarrow h( \frac{x - x_{\rm LC}(t)  }{w(t) } ),
\end{equation}
where we first navigate to the vicinity of the wave front and then probe the possible broadening by the choice of the width function $w(t)$.

\begin{figure}[hbt]
\centering
\subfigure[]{\label{fig:alpha_5_collapse} \includegraphics[width=.48\columnwidth]{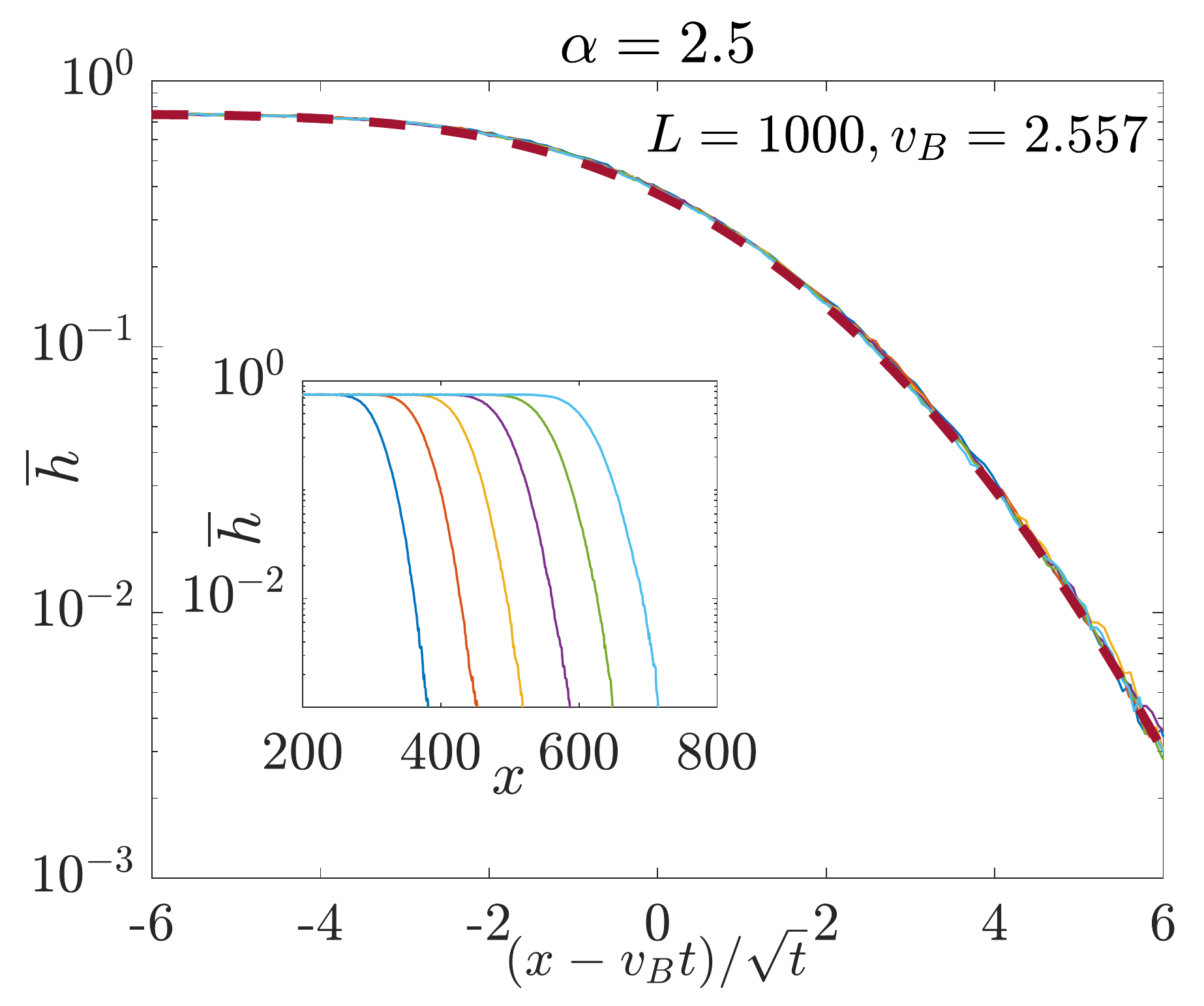}}
\subfigure[]{\label{fig:alpha_35_collapse} \includegraphics[width=.48\columnwidth]{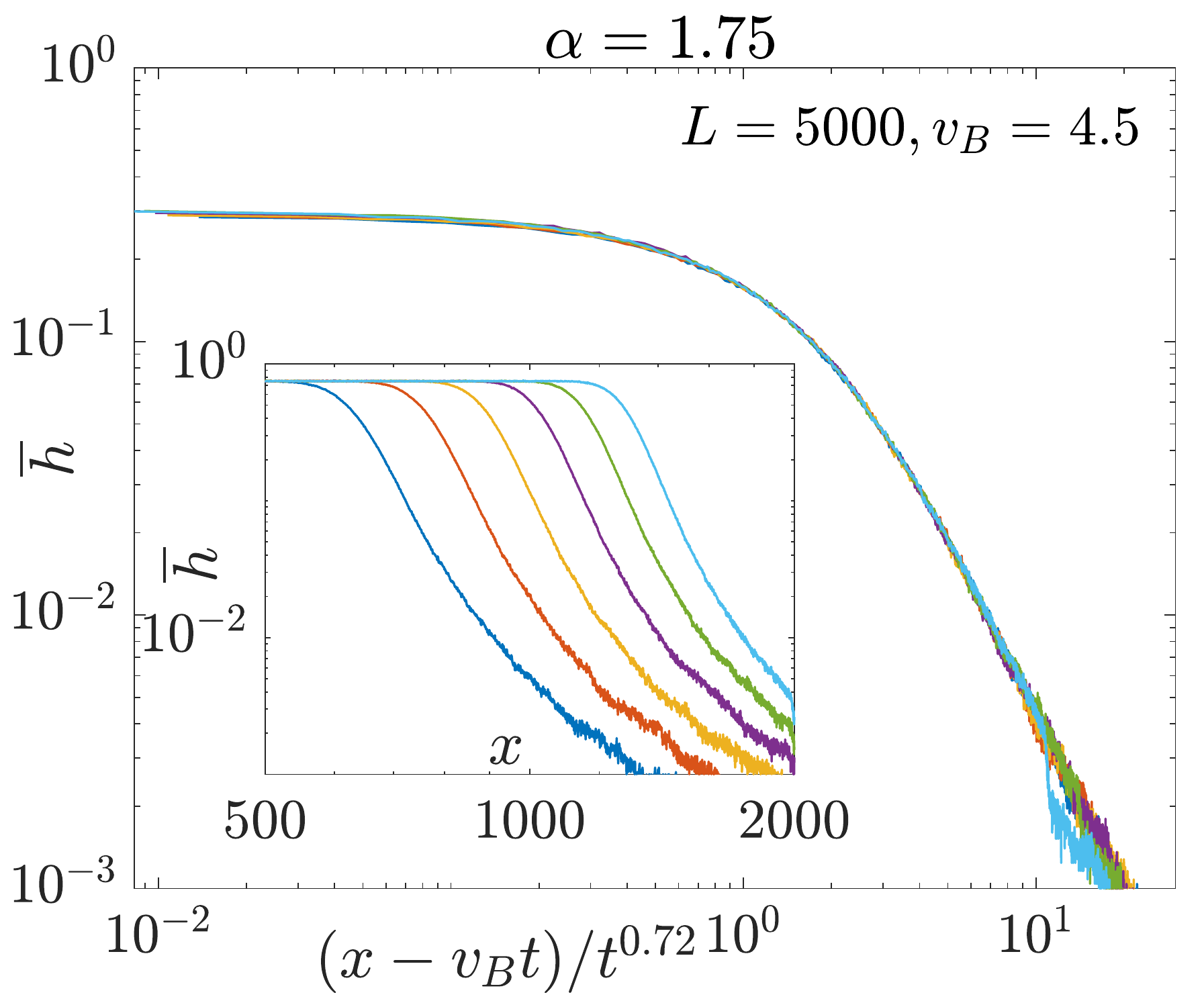}}
\caption{ Data collapse of $\overline{h(x)}$ curves at various $t$ in the linear light cone regime of $\alpha$. Insets show curves before collapsing. (a) The dashed line is the complementary error function $\frac{3}{8} \text{erfc}(x/3.2)$. It matches well with the collapsed curve at $\alpha=2.5$. The curves in the inset have the same time differences. This applies to all the data collapses below. (b) The data collapse of $\overline{h(x)}$ at $\alpha=1.75$. Notice that it has a tail which is close to a power law function.} 
\label{fig:collapse_front_1}
\end{figure}

\begin{figure}[hbt]
\centering
\subfigure[]{\label{fig:alpha_25_collapse} \includegraphics[width=.49\columnwidth]{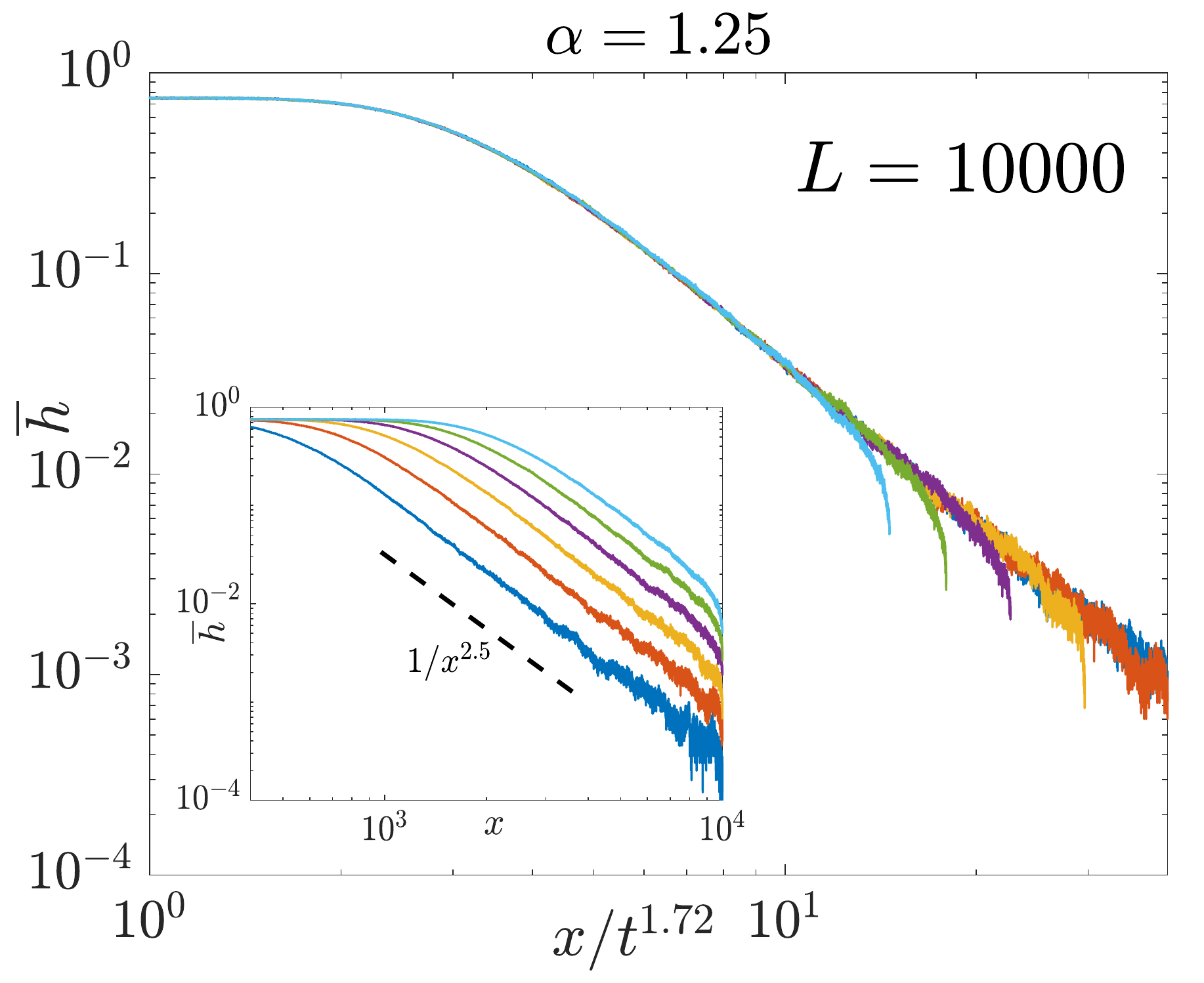}}
\subfigure[]{\label{fig:alpha_15_collapse} \includegraphics[width=.47\columnwidth]{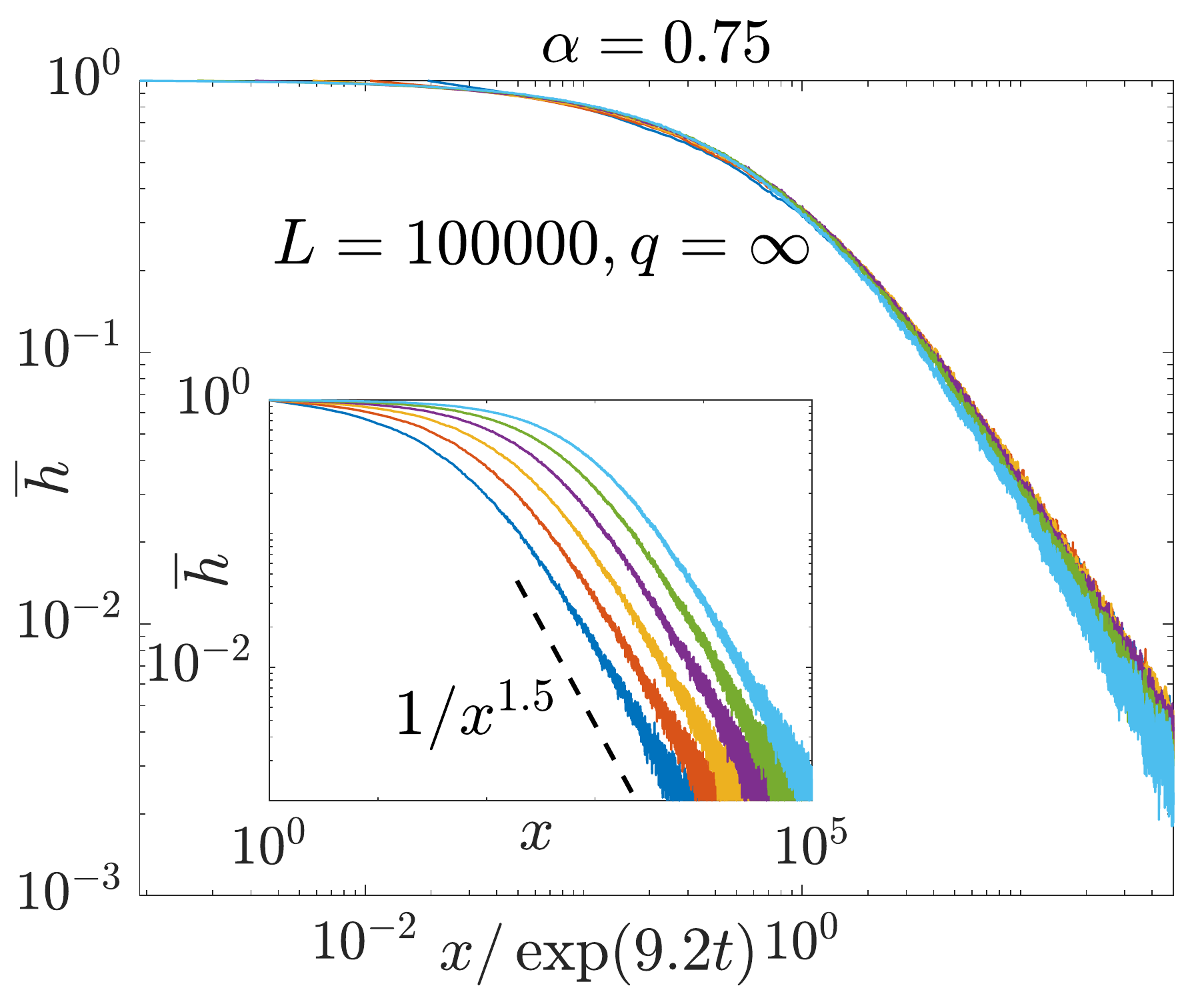}}
\caption{Data collapse of $\overline{h(x)}$ curves for $\alpha=1.25$ (power law light cone) and $\alpha=0.75$ (logarithmic light cone) respectively. In both insets, $\overline{h(x)}$ decays as $x^{ - 2\alpha } $ (dashed line). In (b) we take $q=\infty$ to access larger system size ($L = 100000$).} 
\label{fig:collapse_front_2}
\end{figure}

\begin{figure}[hbt]
\centering
 \subfigure[]{\label{fig:end_point_alpha_5} \includegraphics[width=.48\columnwidth]{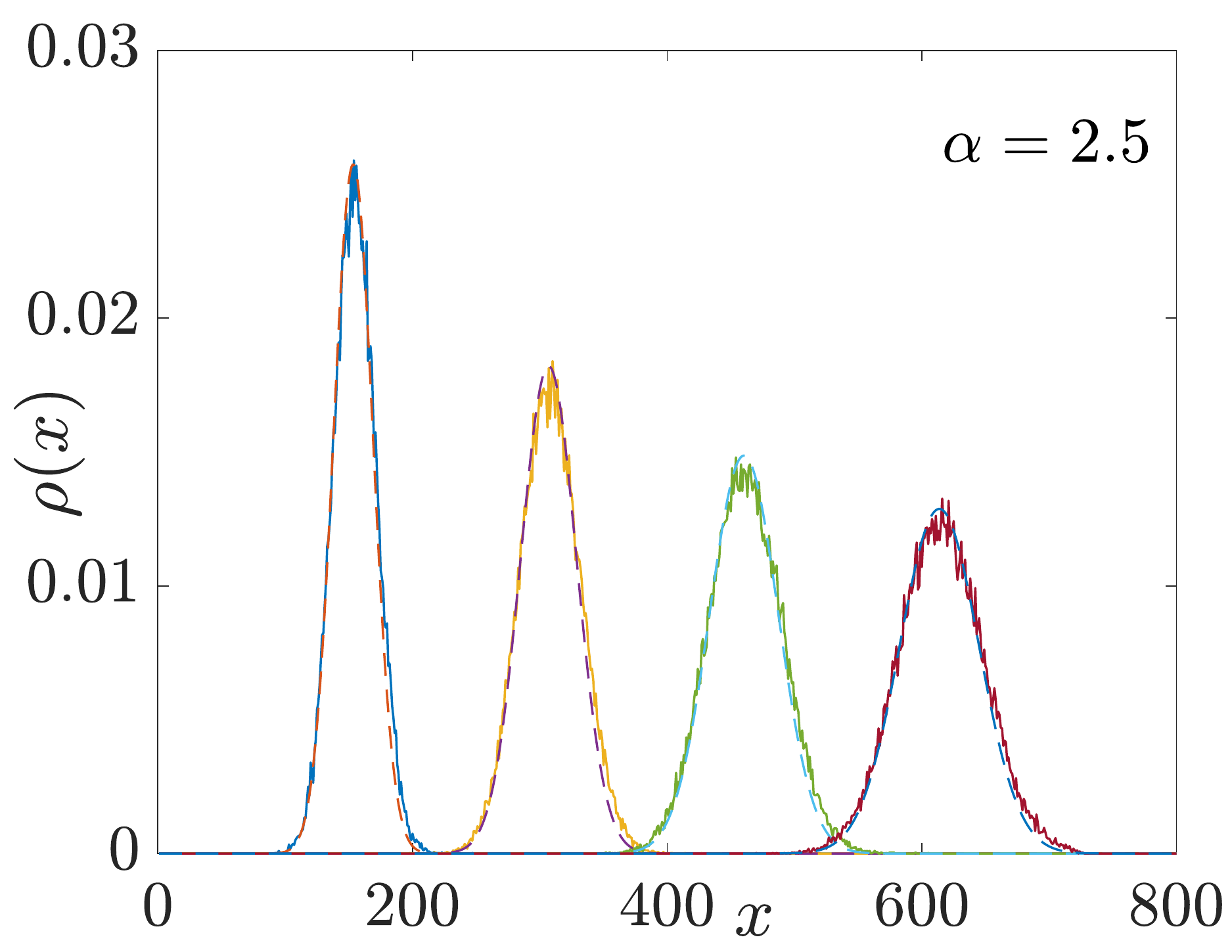}}
  \subfigure[]{\label{fig:end_point_alpha_35} \includegraphics[width=.48\columnwidth]{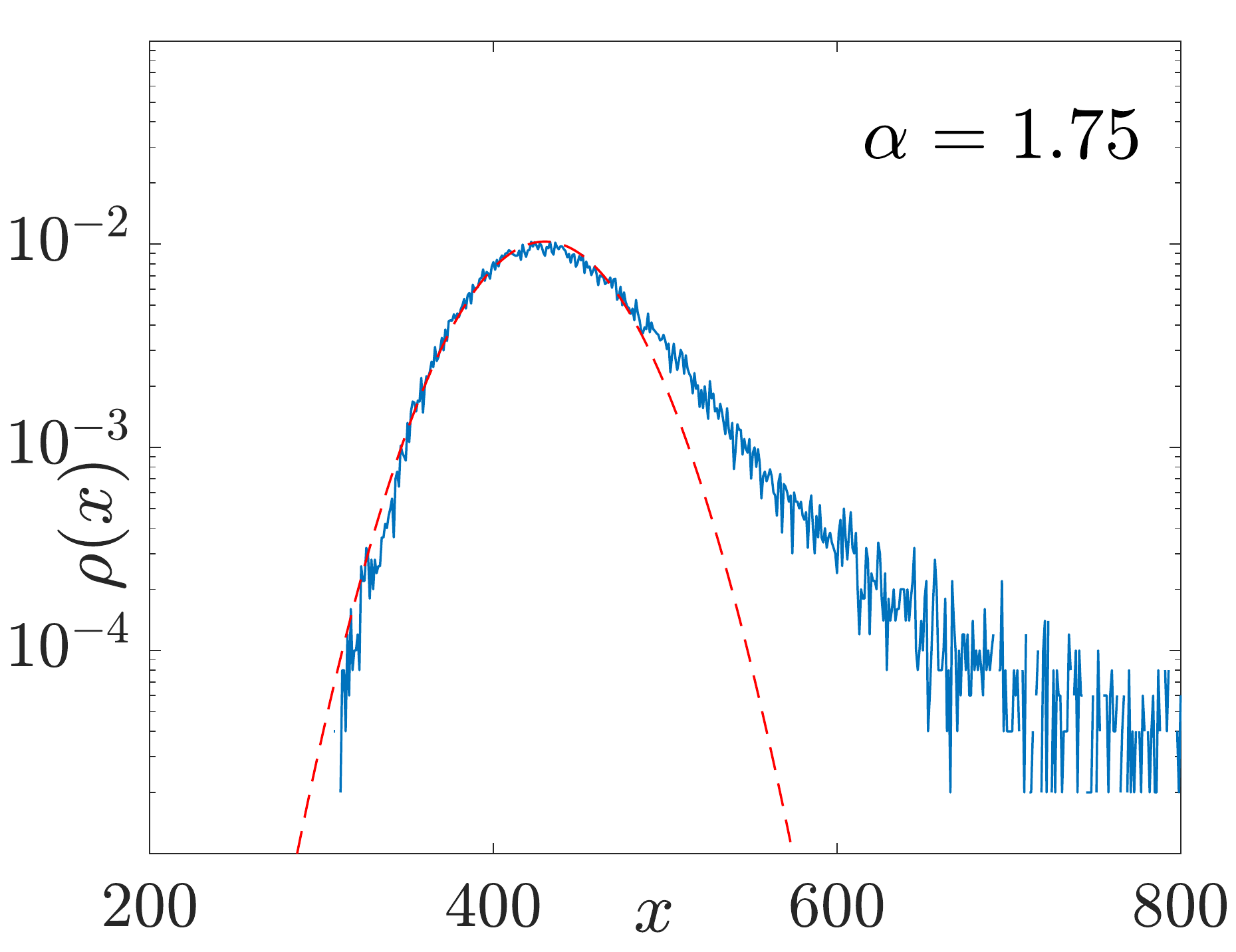}}
\caption{Right end point distribution $\rho(x)$.  (a) When $\alpha=2.5$, $\rho(x)$ at various $t$ (solid curves) fit well with the Gaussian packets which broaden diffusively with time (dashed curves). (b) $\rho(x)$ at $\alpha=1.75$ on the semilog scale (solid curve) compared with Gaussian packet (dashed curve). The deviation is obvious, especially for large $x$. } 
\label{fig:end_point_dis}
\end{figure}

As shown in Fig.~\ref{fig:alpha_5_collapse}, the mean height for $\alpha=2.5$ fits well with the scaling argument $(x-v_Bt)/\sqrt{t}$ with the butterfly velocity $v_B$ as a constant. The front of $\overline{h(x)}$ broadens diffusively as it propagates to the right with $v_B$. Here the front shape of $\overline{h(x,t)}$ can also be determined from the end point distribution $\rho(x)$, which is a Gaussian distribution moving to the right with $v_B$ and broadens diffusively(see Fig.~\ref{fig:end_point_alpha_5}). This means the end point of $h(x)$ is performing a biased random walk. In Fig.~\ref{fig:alpha_5_collapse}, we check that the mean height $\overline{h(x)}$ is equal to the front area of $\rho(y)$, \ie, a complementary error function by $h_{\rm sat}\int_x^{\infty} \rho(y)dy$. We notice similar behaviors for other  values of $\alpha\geq 2$ with $v_B$ decreasing for increasing $\alpha$. \footnote{At $\alpha=2$, it takes very long time for $\rho(x)$ to converge to a Gaussian distribution.} These results provide strong evidence that the physics at $\alpha\geq 2$ is essentially the same as the Brownian circuit with local interaction discussed in Sec.~\ref{sec:master_equ}.

As we  reduce $\alpha$ to the range between about $1.5$ and $2$, although the front still moves with a constant velocity, it broadens super-diffusively with time. We empirically take $w(t) = t^\beta$ with $\ \beta>0.5$. $\beta$ will increase when we decrease $\alpha$ (Fig.~\ref{fig:alpha_35_collapse}). Notice that the tail of the collapsed front is close to a power law decay function. The right side of $\rho(x)$ also has a stretched tail caused by the non-local random walk of the end point(Fig.~\ref{fig:end_point_alpha_35}). But now $\overline{h(x)}$ is not equal to $h_{\rm sat}\int_x^{\infty} \rho(y)dy$. The connection between $\overline{h(x)}$ and $\rho(x)$ will be explored in the future.

When $0.5<\alpha \lesssim 1.5 $, the broadening becomes so wide that $w(t)$ approaches $x_{\rm LC}(t)$. We therefore simply take the scaling argument as $\frac{x}{w(t)}$  or $\frac{x}{x_{\rm LC}(t)}$ and collapse the curves. We further notice that the front decays algebraically in spatial direction with the exponent equal to $2\alpha$ (The inset of Fig.~\ref{fig:alpha_25_collapse} and Fig.~\ref{fig:alpha_15_collapse}). The scaling function is therefore $(\frac{x}{x_{\rm LC}(t)})^{- 2\alpha}$ when $x \gtrsim x_{\rm LC}(t)$. This range of $\alpha$ can be divided into two regimes according to the light cone structure. When $0.8 \lesssim \alpha \lesssim 1.5$, $x_{\rm LC}(t) $ is a power function of time and therefore we have the front (Fig.~\ref{fig:alpha_25_collapse})
\begin{equation}
\overline{h(x,t)}\sim \frac{t^{2 \alpha/\zeta} }{x^{2\alpha}}.
\end{equation}
It is a scale free power function in both spatial and temporal directions. Here $\zeta$ is a function of $\alpha$ which decreases as we reduce $\alpha$. Eventually, when we enter into the logarithmic light cone with $0.5<\alpha \lesssim 0.8$, $x_{\rm LC}(t)  \sim e^{\frac{\lambda}{2\alpha} t}$ and the front of $\overline{h(x,t)}$ approaches a simple scaling form (Fig.~\ref{fig:alpha_15_collapse}),
 \begin{align}
 \overline{h(x,t)}\sim \frac{e^{\lambda t}}{x^{2\alpha}}.
 \end{align}
Therefore we have $\overline{h(x,t)}$ grows exponentially in time with the Lyapunov exponent $\lambda$ as a function $\alpha$.
 
The data collapse results imply the logarithmic, power law and linear light cone formations across different ranges of $\alpha$ and the associated scaling forms of OTOC in the long-range power law interaction, at small $N$ limit. These results are summarized in Table~\ref{tab:N_1}. We expect this feature to be universal and also works for a realistic system with power law interaction.

\begin{table}[htbp]
\centering
\begin{tabular}{c|c|c}
$\alpha$  & Light cone & Scaling form of OTOC \\\hline
$0.5<\alpha \lesssim 0.8$ & Logarithmic & $e^{\lambda t}/x^{2\alpha}$\\\hline
$0.8\lesssim \alpha \lesssim 1.5$  & Power law & $t^{2\alpha/\zeta}/x^{2\alpha}$\\
\hline
$1.5\lesssim \alpha <2$  & Linear & $C((x-v_Bt)/t^\beta)$ \\
\hline
$\alpha\geq 2$  & Linear & $\text{erfc}(a(x-v_Bt)/\sqrt{t})$\\
\end{tabular}
\caption{The light cone structure and the scaling form of OTOC.}
\label{tab:N_1}
\end{table}

\section{Large $N$ limit}
\label{sec:N_infty}

 
In the large $N$ limit, we can directly solve normalized mean height $\underline{h}(x,t)$ from Eq.~\eqref{eq:mean_field}. The equation is similar to the fractional FKPP equation which provides a mean-field description of the reaction-{\it superdiffusion} process\cite{Mancinelli_2002,del-Castillo-Negrete_2003}. It involves two terms: the reaction term and the superdiffusive term determined by the parameter $\alpha$. Ref.~\onlinecite{Mancinelli_2002,del-Castillo-Negrete_2003} show that when $\alpha$ takes proper value, the front of the wave solutions of fractional FKPP equation has the form $\exp( \lambda t  ) / x^{2\alpha}$, which accelerates exponentially in time and decays algebraically in spatial direction. Since $\underline{h}(x,t)$ also ranges from $0$ to $1$, in this section we will also call it $\overline{h(x,t)}$.

We first briefly discuss the two simple limits of Eq.~\eqref{eq:mean_field} at $\alpha=0,\infty$. When $\alpha=0$, the result should match an effective quantum dot with $NL$ spins. In the $\alpha\to\infty$ limit, $D_{ij}$ takes nonzero value only when $i=j\pm 1$. Therefore the evolution equation for $\overline{h(x, t)}$ can be written in the following form
\begin{equation}
\label{eq:mean_field_local}
\begin{aligned}
   \partial_t \overline{h( x, t )}=   D(2\overline{h(x,t)}+\partial_x^2 \overline{h(x,t)}) (1 - \frac{1}{h_{\text{sat}}} \overline{h(x, t )} ).
\end{aligned}
\end{equation}
This equation is very similar to the ordinary  FKPP equation with short-range diffusion term\cite{Fisher1937, kolmogorov1937study}, which provides a mean-field solution for reaction-diffusion process. It has an exponential front $\exp(\lambda (t-x/v_B))$ traveling with constant butterfly velocity $v_B$ without dispersion, contrary to the $N=1$ solution\cite{ablowitz1979, chen_operator_2018, xu2018} whose ballistic wave front broadens as $\sqrt{t}$.

\begin{figure}[hbt]
\centering
 \subfigure[]{\label{fig:linear_log_cone} \includegraphics[width=.48\columnwidth]{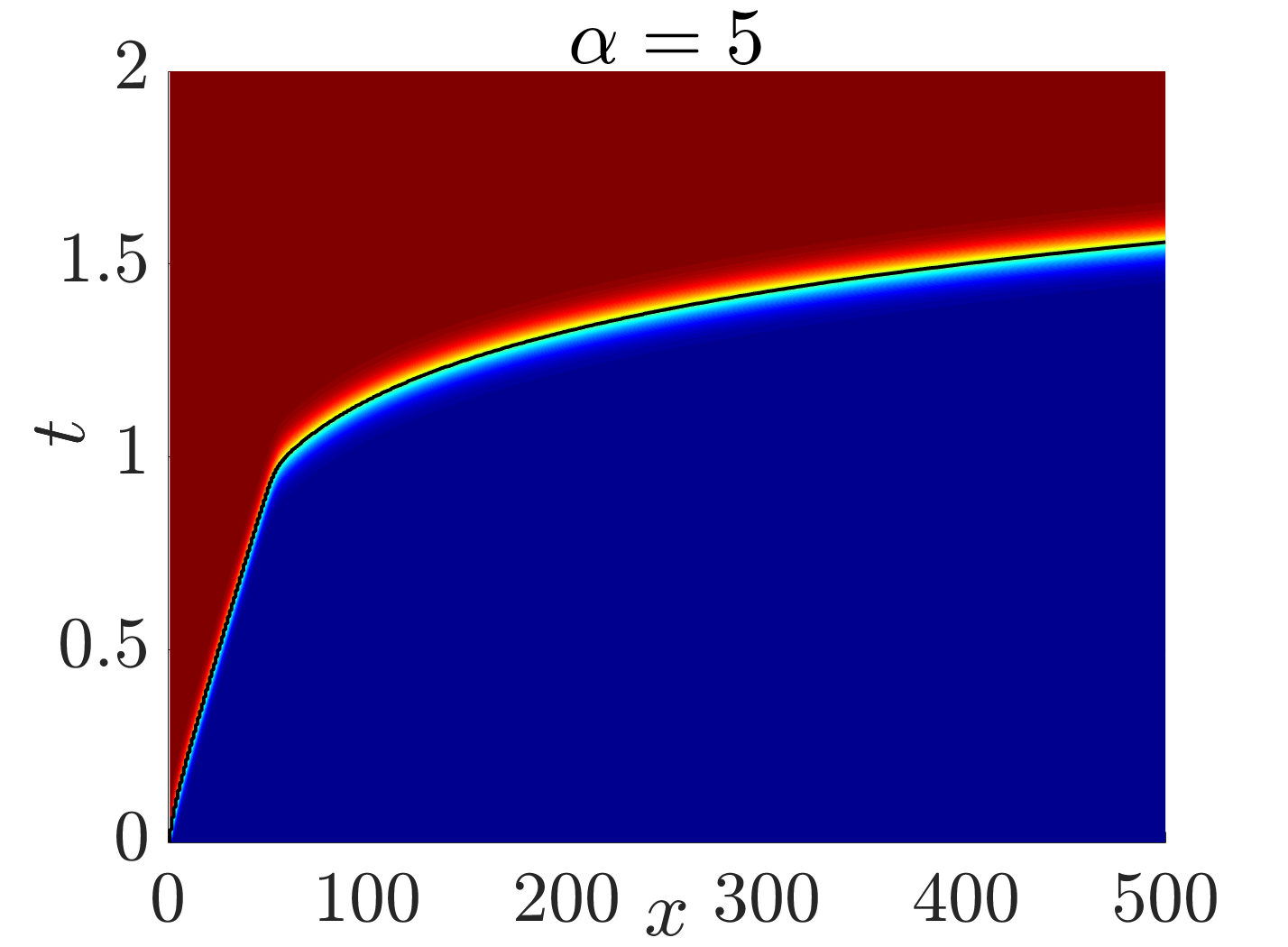}}
  \subfigure[]{\label{fig:light_cone_transition} \includegraphics[width=.48\columnwidth]{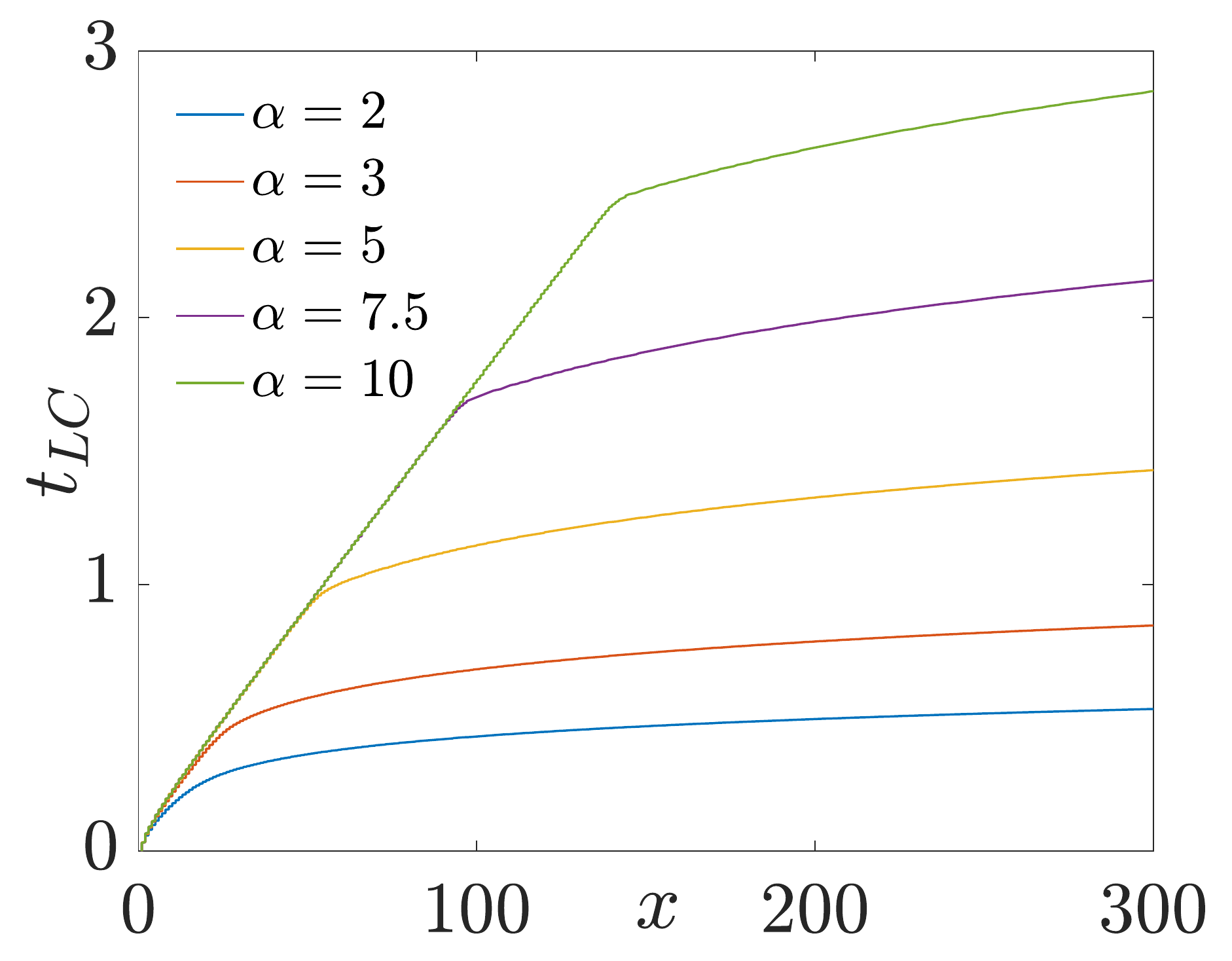}}
\caption{Light cone structures in large $N$ limit. (a) The curve $t_{\rm LC}(x)$ switches from linear to logarithmic at $\alpha=5$. (b) The transitions from linear to logarithmic light cone at various $\alpha$.} 
\label{fig:light_cone_large_N}
\end{figure}

A more striking difference between small $N$ and large $N$ limits appears at finite $\alpha$. In the Sec.~\ref{sec:N_1}, we have shown that in the small $N$ limit, the system has a linear light cone when $\alpha > 1.5$. 
 In contrast, in the large $N$ limit numerical calculation suggests a two-segment structure of the light cone when $\alpha > 1.5$ (Fig.~\ref{fig:light_cone_large_N}): a linear light cone followed by a logarithmic light cone. The latter always appears in the late time. More quantitatively, we find three interesting phenomena (see Fig.~\ref{fig:light_cone_transition}):
\begin{enumerate}
\item The butterfly velocity in the linear light cone regime is independent of $\alpha$. 
\item The logarithmic light cone scales as $t_{\rm LC}\sim \alpha\log x$.
\item The transition from linear light cone to logarithmic light cone occurs at the intersection $x \sim \alpha \log x $.
\end{enumerate}

\begin{figure}[hbt]
\centering
 \subfigure[]{\label{fig:linear_cone_large_N} \includegraphics[width=.48\columnwidth]{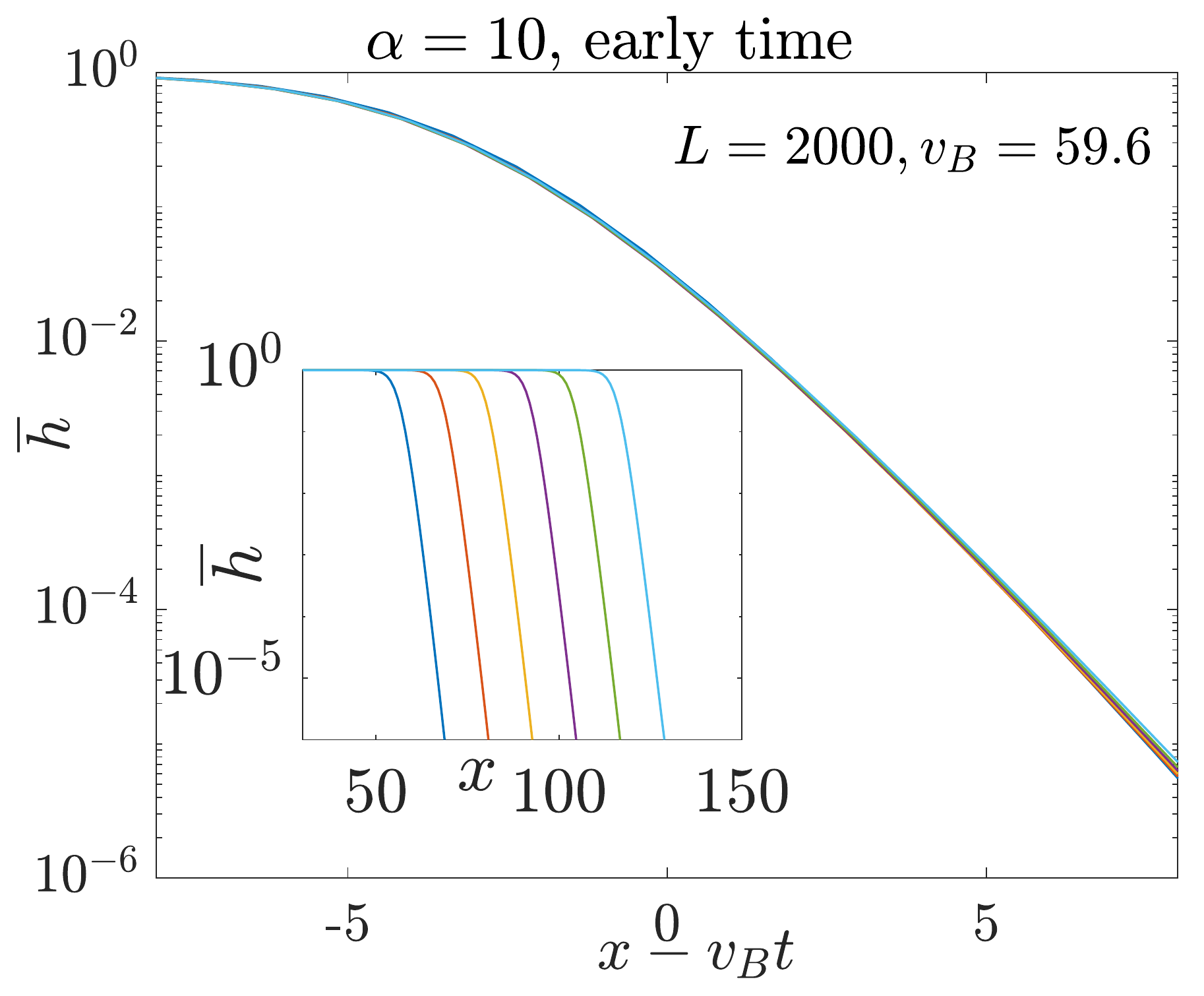}}
  \subfigure[]{\label{fig:log_cone_large_N} \includegraphics[width=.47\columnwidth]{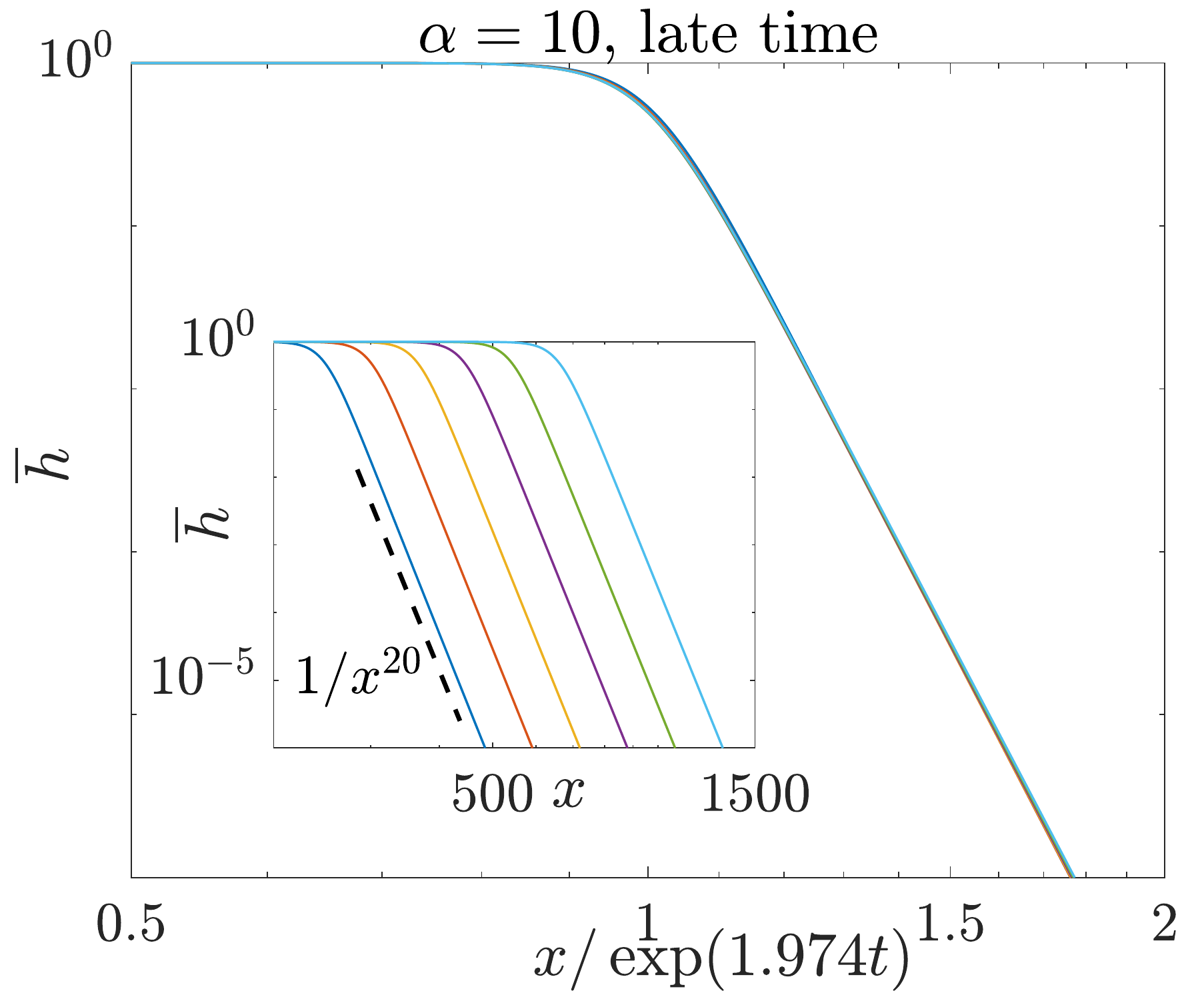}}
\caption{The scaling of the front in the linear-to-logarithmic transition in the large $N$ limit at $\alpha=10$. (a) Scaling collapse for the exponential tail on the semi-log scale at early time in the linear light cone regime. The front propagates linearly in time. (b) Scaling collapse on the log-log scale at late time in the logarithmic light cone regime. The front propagates exponentially in time.  The dashed linear scales as $1/x^{20}$. } 
\label{fig:collapse_large_N}
\end{figure}

\begin{figure}[hbt]
\centering
\subfigure[]{\label{fig:alpha_20_front} \includegraphics[width=.48\columnwidth]{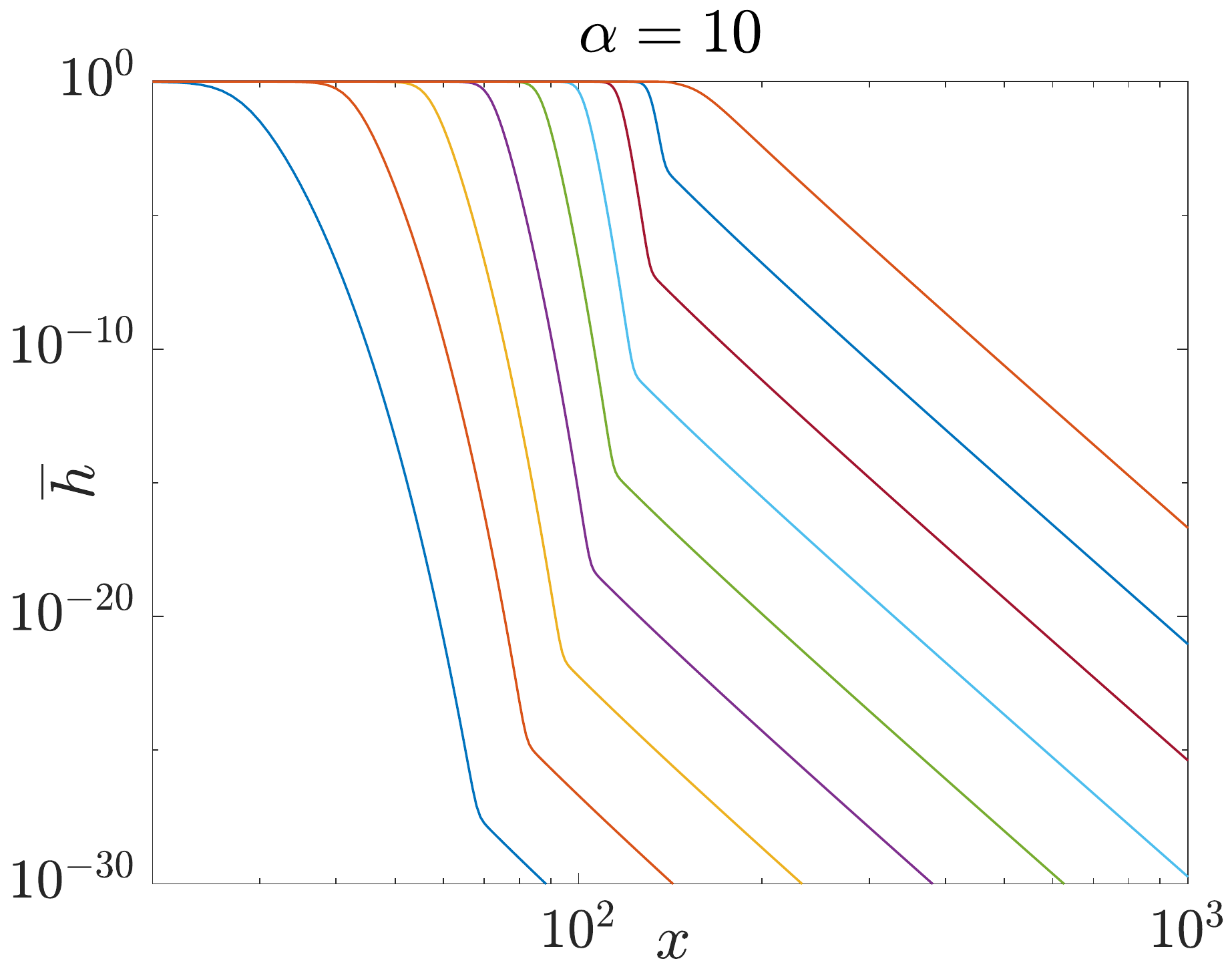}}
 \subfigure[]{\label{fig:log_cone_large_N_alpha_2} \includegraphics[width=.47\columnwidth]{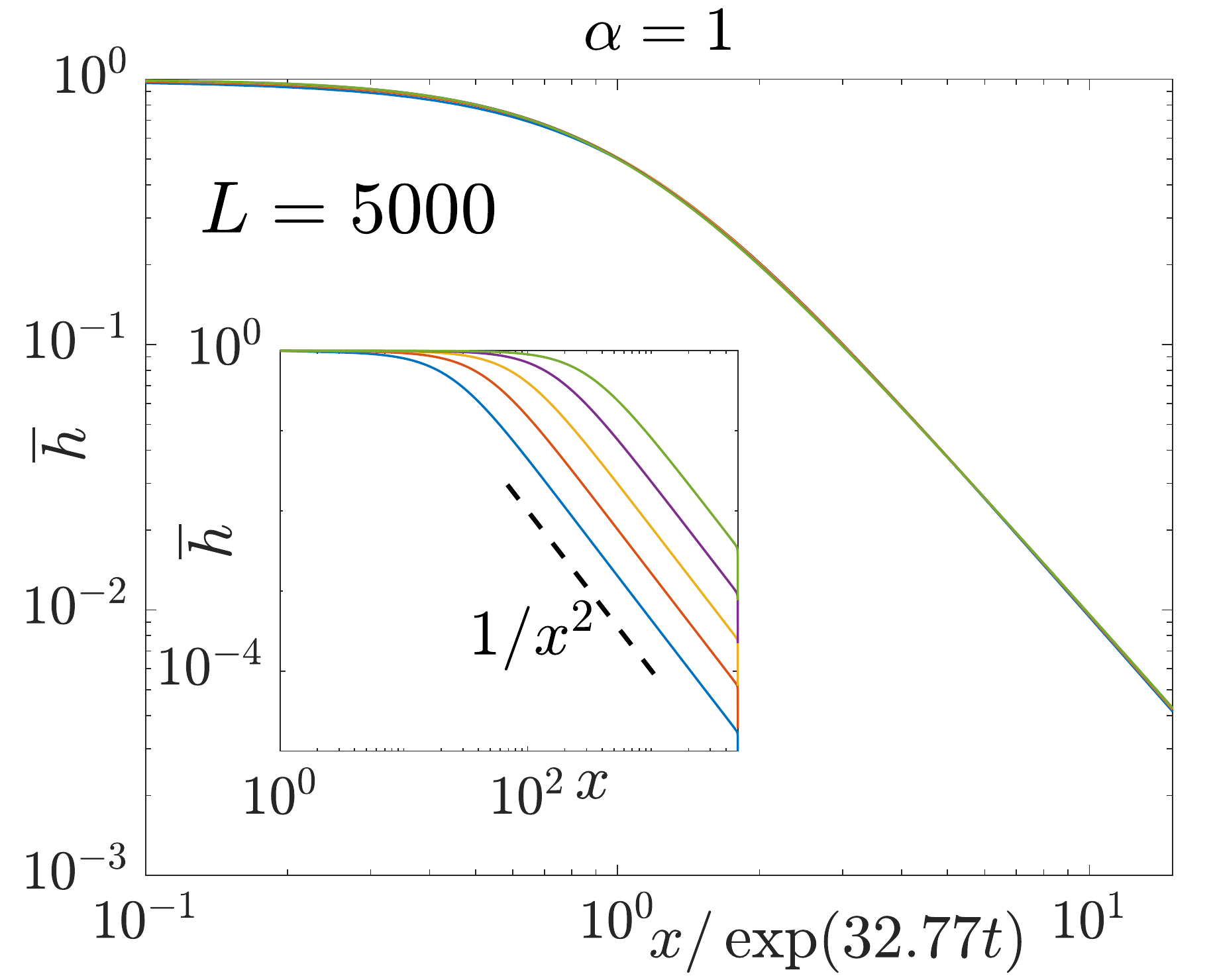}}
\caption{ (a) The front shape of $\overline{h(x,t )}$ vs $x$ at different times on the semi-log scale for $\alpha=10$. (b) Scaling collapse of $\overline{h(x, t)}$ on the log-log scale at $\alpha=1$. The dashed linear scales as $1/x^{2}$. } 
\end{figure}

The early time linear light cone is easy to understand.  When $\alpha$ is sufficiently large, the power law kernel in Eq.~\eqref{eq:mean_field} is numerically close to the diffusive kernel and therefore at early times Eq.~\eqref{eq:mean_field} describes the ordinary FKPP dynamics with the front scaling as $\exp(\lambda(t-x/v_B))$ (see the data collapse in Fig.~\ref{fig:linear_cone_large_N}). Far ahead of this exponential front, we further notice a power law tail $x^{-2\alpha}$ as shown in Fig.~\ref{fig:alpha_20_front}. It is initially buried far away from the front and has comparable order with the exponential front beyond the location determined by $\exp(-\lambda x/v_B) \sim x^{-2\alpha}$.  When the front reaches this location, the power law tail becomes dominant and destroys the linear light cone.  
Indeed, in the data collapse for the late time front dynamics at $\alpha=10$ (Fig.~\ref{fig:log_cone_large_N}), we see the front scales as a power law function $e^{\lambda t}/x^{2\alpha}$ at late times.
Similar linear followed by logarithmic light cone behavior has also been discussed in Ref.~\onlinecite{Coulon_2012, Mancinelli_2002, del-Castillo-Negrete_2003, Gong2014}. 
When $\alpha < 1.5$, only logarithmic light cone appears with the front scaling $e^{\lambda t}/x^{2\alpha}$ (see Fig.~\ref{fig:log_cone_large_N_alpha_2}).
Although the asymptotic logarithmic light cone appears counter-intuitively fast, it does not violate the information bound of the power law interaction systems proposed in Ref.~\onlinecite{Gong2014,Foss-Feig2015,Storch_2015,matsuta_improving_2017,Tran2018,else_improved_2018}, which only works for systems with finite $N$.

Based on the analysis above, we find that $N=1$ and $N=\infty$ give different light cone structures. The large $N$ solution (fraction FKPP equation) can be understood as the mean field approximation of the master equation. When $\alpha < 0.5$, almost all the sites are coupled together, thus justifying the large $N$ mean field limit. When $\alpha > 0.5 $, the mean field approximation starts to break down. The large $N$ solution does not have a power law light cone regime as in the case of $N = 1$. When $\alpha > 1.5$, we also see that the initial linear light cone is tamed by the power law tail of the front in the long time.


The crossover for the dynamics from finite $N$ to large $N$ is an intriguing question. As discussed in Ref.\ \onlinecite{Brunet_1997, Kessler_1998, Panja_2004}, the large $N$ solution in diffusion-reaction process is unstable against the $1/N$ correction, which can be effectively treated as a noise term in FKPP equation. Asymptotically, the noise generates  fluctuation that leads to a diffusively broadening of the  otherwise dispersion-less wavefront. The authors of Ref.~\onlinecite{xu2018} applied this result to discuss this crossover in chaotic systems with local interaction and showed that in the long time limit, the chaos dynamics at large but finite $N$ is qualitatively the same as the small $N$ limit.

We expect the $1/N$ correction could also significantly change spatiotemporal behavior of chaos dynamics in the power law interaction systems. As long as $N$ is finite, the large $N$ solution for the front is not stable. When $\alpha$ is large, we should only observe the linear rather than the  two-segment light cone behavior. Indeed, the asymptotic linear light cone structure has been found in some reaction-superdiffusion kinetics with large but finite $N$\cite{Brockmann_2007}. Moreover, a power law light cone at intermediate $\alpha$ could also appear, which is caused by the competition between $1/N$ effect and non-local hopping process. A good starting point to explore the profound $1/N$ effect is to introduce noise term in Eq.~\eqref{eq:mean_field}, which we leave for future work.

\section{Conclusion and Outlook}
\label{sec:conclusion}

In conclusion, we use OTOC $C(x, t)$ to diagnose the chaos propagation in one dimensional power law interaction systems with $N$ number of qubits at each site. In the $N=1$ limit, by using $C(x,t)$ to define light cone, we find (1) a logarithmic light cone regime when $0.5<\alpha \lesssim 0.8$,  (2) a power law light cone regime when $0.8 \lesssim \alpha \lesssim 1.5$ and (3) an emergent linear light cone regime when $\alpha \gtrsim 1.5$. The linear light cone regime can be further divided into two sub-regimes according to the different scaling behaviors of $C(x,t)$. When $\alpha\geq 2$, the front of $C(x,t)$ is broadened diffusively in time, the same as systems with local interaction. This result suggests in this regime, the locality can be fully recovered in the long-range power law interaction systems. When $1.5 \lesssim \alpha<2$, the front of $C(x,t)$ is still moving with constant butterfly velocity but broadens superdiffusively in time. In the power law light cone regime, we find that the front of $C(x,t)\sim t^{2 \alpha / \zeta }/x^{2\alpha}$, \ie, an algebraic dependence in both temporal and spatial directions. In the logarithmic light cone regime, the above scaling function is replaced by $C(t,x)\sim \exp(\lambda t)/x^{2\alpha}$. Finally, when $\alpha<0.5$, the locality is completely lost and $C(x, t)$ shows similar behavior as in the $\alpha=0$ limit.

We also investigate the scaling function of OTOC in the large $N$ limit.  Besides the pure logarithmic light with $\alpha \lesssim 1.5 $, we also find a transition from early time linear light cone to late time logarithmic light cone behavior when $\alpha \gtrsim 1.5$. We comment on the stability of this late time logarithmic light cone behavior at finite $N$ and argue that the large $N$ solution should be unstable against $1/N$ correction. 


Our work opens the door to a number of intriguing future directions. First, our analysis on the small $N$ limit in one dimension can be extended to system with finite $N$ or higher dimensions and therefore gives a complete physical picture for chaos in long range power law interaction system. Moreover, our result suggests possible improvement for optimizing the information bound in long range interaction system. Another interesting direction would be to understand other dynamically related quantities, such as the entanglement growth and thermalization rate in a generic chaotic systems with long range interaction.


\acknowledgements

We acknowledge useful discussion with  Sarang Gopalakrishnan, Alexey Gorshkov, Rajibul Islam, Austen Lamacraft, Andreas W.W. Ludwig,  Marcos Rigol, Shinsei Ryu, Brian Swingle, Minh Tran, Cenke Xu and Shenglong Xu. We also thank the accommodation and interactive environment of the KITP program ``The Dynamics of Quantum Information". XC and TZ are supported by postdoctoral fellowships from the Gordon and Betty Moore Foundation, under the EPiQS initiative, Grant GBMF4304, at the Kavli Institute for Theoretical Physics. This research was supported in part by the National Science Foundation under Grant No. NSF PHY-1748958.
We acknowledge support from the Center for Scientific Computing from the CNSI, MRL: an NSF MRSEC (DMR-1720256) and NSF CNS-1725797.

\bibliographystyle{apsrev4-1}
\bibliography{pow_law}

\begin{thebibliography}{46}%
\makeatletter
\providecommand \@ifxundefined [1]{%
 \@ifx{#1\undefined}
}%
\providecommand \@ifnum [1]{%
 \ifnum #1\expandafter \@firstoftwo
 \else \expandafter \@secondoftwo
 \fi
}%
\providecommand \@ifx [1]{%
 \ifx #1\expandafter \@firstoftwo
 \else \expandafter \@secondoftwo
 \fi
}%
\providecommand \natexlab [1]{#1}%
\providecommand \enquote  [1]{``#1''}%
\providecommand \bibnamefont  [1]{#1}%
\providecommand \bibfnamefont [1]{#1}%
\providecommand \citenamefont [1]{#1}%
\providecommand \href@noop [0]{\@secondoftwo}%
\providecommand \href [0]{\begingroup \@sanitize@url \@href}%
\providecommand \@href[1]{\@@startlink{#1}\@@href}%
\providecommand \@@href[1]{\endgroup#1\@@endlink}%
\providecommand \@sanitize@url [0]{\catcode `\\12\catcode `\$12\catcode
  `\&12\catcode `\#12\catcode `\^12\catcode `\_12\catcode `\%12\relax}%
\providecommand \@@startlink[1]{}%
\providecommand \@@endlink[0]{}%
\providecommand \url  [0]{\begingroup\@sanitize@url \@url }%
\providecommand \@url [1]{\endgroup\@href {#1}{\urlprefix }}%
\providecommand \urlprefix  [0]{URL }%
\providecommand \Eprint [0]{\href }%
\providecommand \doibase [0]{http://dx.doi.org/}%
\providecommand \selectlanguage [0]{\@gobble}%
\providecommand \bibinfo  [0]{\@secondoftwo}%
\providecommand \bibfield  [0]{\@secondoftwo}%
\providecommand \translation [1]{[#1]}%
\providecommand \BibitemOpen [0]{}%
\providecommand \bibitemStop [0]{}%
\providecommand \bibitemNoStop [0]{.\EOS\space}%
\providecommand \EOS [0]{\spacefactor3000\relax}%
\providecommand \BibitemShut  [1]{\csname bibitem#1\endcsname}%
\let\auto@bib@innerbib\@empty
\bibitem [{\citenamefont {Larkin}\ and\ \citenamefont
  {Ovchinnikov}(1969)}]{larkin1969}%
  \BibitemOpen
  \bibfield  {author} {\bibinfo {author} {\bibfnamefont {A.~I.}\ \bibnamefont
  {Larkin}}\ and\ \bibinfo {author} {\bibfnamefont {Y.~N.}\ \bibnamefont
  {Ovchinnikov}},\ }\href {http://adsabs.harvard.edu/abs/1969JETP...28.1200L}
  {\bibfield  {journal} {\bibinfo  {journal} {Soviet Journal of Experimental
  and Theoretical Physics}\ }\textbf {\bibinfo {volume} {28}},\ \bibinfo
  {pages} {1200} (\bibinfo {year} {1969})}\BibitemShut {NoStop}%
\bibitem [{\citenamefont {Hayden}\ and\ \citenamefont
  {Preskill}(2007)}]{hayden2007}%
  \BibitemOpen
  \bibfield  {author} {\bibinfo {author} {\bibfnamefont {P.}~\bibnamefont
  {Hayden}}\ and\ \bibinfo {author} {\bibfnamefont {J.}~\bibnamefont
  {Preskill}},\ }\href {http://stacks.iop.org/1126-6708/2007/i=09/a=120}
  {\bibfield  {journal} {\bibinfo  {journal} {Journal of High Energy Physics}\
  }\textbf {\bibinfo {volume} {2007}},\ \bibinfo {pages} {120} (\bibinfo {year}
  {2007})}\BibitemShut {NoStop}%
\bibitem [{\citenamefont {Sekino}\ and\ \citenamefont
  {Susskind}(2008)}]{sekino2008}%
  \BibitemOpen
  \bibfield  {author} {\bibinfo {author} {\bibfnamefont {Y.}~\bibnamefont
  {Sekino}}\ and\ \bibinfo {author} {\bibfnamefont {L.}~\bibnamefont
  {Susskind}},\ }\href {http://stacks.iop.org/1126-6708/2008/i=10/a=065}
  {\bibfield  {journal} {\bibinfo  {journal} {Journal of High Energy Physics}\
  }\textbf {\bibinfo {volume} {2008}},\ \bibinfo {pages} {065} (\bibinfo {year}
  {2008})}\BibitemShut {NoStop}%
\bibitem [{\citenamefont {Kitaev}(2015)}]{kitaev2015}%
  \BibitemOpen
  \bibfield  {author} {\bibinfo {author} {\bibfnamefont {A.}~\bibnamefont
  {Kitaev}},\ }\href@noop {} {} (\bibinfo {year} {2015}),\ \bibinfo {note}
  {talks at KITP, April 7, 2015 and May 27, 2015}\BibitemShut {NoStop}%
\bibitem [{\citenamefont {{Roberts}}\ \emph {et~al.}(2015)\citenamefont
  {{Roberts}}, \citenamefont {{Stanford}},\ and\ \citenamefont
  {{Susskind}}}]{roberts2015b}%
  \BibitemOpen
  \bibfield  {author} {\bibinfo {author} {\bibfnamefont {D.~A.}\ \bibnamefont
  {{Roberts}}}, \bibinfo {author} {\bibfnamefont {D.}~\bibnamefont
  {{Stanford}}}, \ and\ \bibinfo {author} {\bibfnamefont {L.}~\bibnamefont
  {{Susskind}}},\ }\href {\doibase 10.1007/JHEP03(2015)051} {\bibfield
  {journal} {\bibinfo  {journal} {Journal of High Energy Physics}\ }\textbf
  {\bibinfo {volume} {3}},\ \bibinfo {eid} {51} (\bibinfo {year} {2015})},\
  \bibinfo {note} {arXiv: 1409.8180}\BibitemShut {NoStop}%
\bibitem [{\citenamefont {Maldacena}\ and\ \citenamefont
  {Stanford}(2016)}]{maldacena2016}%
  \BibitemOpen
  \bibfield  {author} {\bibinfo {author} {\bibfnamefont {J.}~\bibnamefont
  {Maldacena}}\ and\ \bibinfo {author} {\bibfnamefont {D.}~\bibnamefont
  {Stanford}},\ }\href {\doibase 10.1103/PhysRevD.94.106002} {\bibfield
  {journal} {\bibinfo  {journal} {Phys. Rev. D}\ }\textbf {\bibinfo {volume}
  {94}},\ \bibinfo {pages} {106002} (\bibinfo {year} {2016})}\BibitemShut
  {NoStop}%
\bibitem [{\citenamefont {Shenker}\ and\ \citenamefont
  {Stanford}(2014)}]{shenker2013a}%
  \BibitemOpen
  \bibfield  {author} {\bibinfo {author} {\bibfnamefont {S.~H.}\ \bibnamefont
  {Shenker}}\ and\ \bibinfo {author} {\bibfnamefont {D.}~\bibnamefont
  {Stanford}},\ }\href {\doibase 10.1007/JHEP03(2014)067} {\bibfield  {journal}
  {\bibinfo  {journal} {Journal of High Energy Physics}\ }\textbf {\bibinfo
  {volume} {2014}},\ \bibinfo {pages} {1} (\bibinfo {year} {2014})}\BibitemShut
  {NoStop}%
\bibitem [{\citenamefont {Roberts}\ and\ \citenamefont
  {Stanford}(2015)}]{Roberts2015}%
  \BibitemOpen
  \bibfield  {author} {\bibinfo {author} {\bibfnamefont {D.~A.}\ \bibnamefont
  {Roberts}}\ and\ \bibinfo {author} {\bibfnamefont {D.}~\bibnamefont
  {Stanford}},\ }\href {\doibase 10.1103/PhysRevLett.115.131603} {\bibfield
  {journal} {\bibinfo  {journal} {Phys. Rev. Lett.}\ }\textbf {\bibinfo
  {volume} {115}},\ \bibinfo {pages} {131603} (\bibinfo {year}
  {2015})}\BibitemShut {NoStop}%
\bibitem [{\citenamefont {Gu}\ \emph {et~al.}(2017)\citenamefont {Gu},
  \citenamefont {Qi},\ and\ \citenamefont {Stanford}}]{gu2017}%
  \BibitemOpen
  \bibfield  {author} {\bibinfo {author} {\bibfnamefont {Y.}~\bibnamefont
  {Gu}}, \bibinfo {author} {\bibfnamefont {X.-L.}\ \bibnamefont {Qi}}, \ and\
  \bibinfo {author} {\bibfnamefont {D.}~\bibnamefont {Stanford}},\ }\href
  {\doibase 10.1007/JHEP05(2017)125} {\bibfield  {journal} {\bibinfo  {journal}
  {Journal of High Energy Physics}\ }\textbf {\bibinfo {volume} {2017}},\
  \bibinfo {pages} {125} (\bibinfo {year} {2017})}\BibitemShut {NoStop}%
\bibitem [{\citenamefont {von Keyserlingk}\ \emph {et~al.}(2018)\citenamefont
  {von Keyserlingk}, \citenamefont {Rakovszky}, \citenamefont {Pollmann},\ and\
  \citenamefont {Sondhi}}]{keyserlingk2017}%
  \BibitemOpen
  \bibfield  {author} {\bibinfo {author} {\bibfnamefont {C.~W.}\ \bibnamefont
  {von Keyserlingk}}, \bibinfo {author} {\bibfnamefont {T.}~\bibnamefont
  {Rakovszky}}, \bibinfo {author} {\bibfnamefont {F.}~\bibnamefont {Pollmann}},
  \ and\ \bibinfo {author} {\bibfnamefont {S.~L.}\ \bibnamefont {Sondhi}},\
  }\href {\doibase 10.1103/PhysRevX.8.021013} {\bibfield  {journal} {\bibinfo
  {journal} {Phys. Rev. X}\ }\textbf {\bibinfo {volume} {8}},\ \bibinfo {pages}
  {021013} (\bibinfo {year} {2018})}\BibitemShut {NoStop}%
\bibitem [{\citenamefont {Nahum}\ \emph {et~al.}(2018)\citenamefont {Nahum},
  \citenamefont {Vijay},\ and\ \citenamefont {Haah}}]{nahum2017}%
  \BibitemOpen
  \bibfield  {author} {\bibinfo {author} {\bibfnamefont {A.}~\bibnamefont
  {Nahum}}, \bibinfo {author} {\bibfnamefont {S.}~\bibnamefont {Vijay}}, \ and\
  \bibinfo {author} {\bibfnamefont {J.}~\bibnamefont {Haah}},\ }\href {\doibase
  10.1103/PhysRevX.8.021014} {\bibfield  {journal} {\bibinfo  {journal} {Phys.
  Rev. X}\ }\textbf {\bibinfo {volume} {8}},\ \bibinfo {pages} {021014}
  (\bibinfo {year} {2018})}\BibitemShut {NoStop}%
\bibitem [{\citenamefont {Leviatan}\ \emph {et~al.}(2017)\citenamefont
  {Leviatan}, \citenamefont {Pollmann}, \citenamefont {Bardarson},
  \citenamefont {Huse},\ and\ \citenamefont {Altman}}]{leviatan2017}%
  \BibitemOpen
  \bibfield  {author} {\bibinfo {author} {\bibfnamefont {E.}~\bibnamefont
  {Leviatan}}, \bibinfo {author} {\bibfnamefont {F.}~\bibnamefont {Pollmann}},
  \bibinfo {author} {\bibfnamefont {J.~H.}\ \bibnamefont {Bardarson}}, \bibinfo
  {author} {\bibfnamefont {D.~A.}\ \bibnamefont {Huse}}, \ and\ \bibinfo
  {author} {\bibfnamefont {E.}~\bibnamefont {Altman}},\ }\href
  {http://arxiv.org/abs/1702.08894} {\bibfield  {journal} {\bibinfo  {journal}
  {arXiv:1702.08894 [cond-mat, physics:quant-ph]}\ } (\bibinfo {year}
  {2017})},\ \bibinfo {note} {arXiv: 1702.08894}\BibitemShut {NoStop}%
\bibitem [{\citenamefont {{Xu}}\ and\ \citenamefont
  {{Swingle}}(2018)}]{xu2018}%
  \BibitemOpen
  \bibfield  {author} {\bibinfo {author} {\bibfnamefont {S.}~\bibnamefont
  {{Xu}}}\ and\ \bibinfo {author} {\bibfnamefont {B.}~\bibnamefont
  {{Swingle}}},\ }\href@noop {} {\bibfield  {journal} {\bibinfo  {journal}
  {ArXiv e-prints}\ } (\bibinfo {year} {2018})},\ \bibinfo {note} {arXiv:
  1802.00801}\BibitemShut {NoStop}%
\bibitem [{\citenamefont {Lieb}\ and\ \citenamefont
  {Robinson}(1972)}]{lieb1972}%
  \BibitemOpen
  \bibfield  {author} {\bibinfo {author} {\bibfnamefont {E.~H.}\ \bibnamefont
  {Lieb}}\ and\ \bibinfo {author} {\bibfnamefont {D.~W.}\ \bibnamefont
  {Robinson}},\ }\href {\doibase 10.1007/BF01645779} {\bibfield  {journal}
  {\bibinfo  {journal} {Communications in Mathematical Physics}\ }\textbf
  {\bibinfo {volume} {28}},\ \bibinfo {pages} {251} (\bibinfo {year}
  {1972})}\BibitemShut {NoStop}%
\bibitem [{\citenamefont {{Blatt}}\ and\ \citenamefont
  {{Roos}}(2012)}]{Blatt_2012}%
  \BibitemOpen
  \bibfield  {author} {\bibinfo {author} {\bibfnamefont {R.}~\bibnamefont
  {{Blatt}}}\ and\ \bibinfo {author} {\bibfnamefont {C.~F.}\ \bibnamefont
  {{Roos}}},\ }\href {\doibase 10.1038/nphys2252} {\bibfield  {journal}
  {\bibinfo  {journal} {Nature Physics}\ }\textbf {\bibinfo {volume} {8}},\
  \bibinfo {pages} {277} (\bibinfo {year} {2012})}\BibitemShut {NoStop}%
\bibitem [{\citenamefont {{Bloch}}\ \emph {et~al.}(2012)\citenamefont
  {{Bloch}}, \citenamefont {{Dalibard}},\ and\ \citenamefont
  {S.}}]{Bloch_2012}%
  \BibitemOpen
  \bibfield  {author} {\bibinfo {author} {\bibfnamefont {I.}~\bibnamefont
  {{Bloch}}}, \bibinfo {author} {\bibfnamefont {J.}~\bibnamefont {{Dalibard}}},
  \ and\ \bibinfo {author} {\bibfnamefont {N.}~\bibnamefont {S.}},\ }\href
  {\doibase doi.org/10.1038/nphys2259} {\bibfield  {journal} {\bibinfo
  {journal} {Nature Physics}\ }\textbf {\bibinfo {volume} {8}},\ \bibinfo
  {pages} {267} (\bibinfo {year} {2012})}\BibitemShut {NoStop}%
\bibitem [{\citenamefont {{Awschalom}}\ \emph {et~al.}(2018)\citenamefont
  {{Awschalom}}, \citenamefont {{Hanson}}, \citenamefont {J.},\ and\
  \citenamefont {B.}}]{Awschalom_2012}%
  \BibitemOpen
  \bibfield  {author} {\bibinfo {author} {\bibfnamefont {D.}~\bibnamefont
  {{Awschalom}}}, \bibinfo {author} {\bibfnamefont {R.}~\bibnamefont
  {{Hanson}}}, \bibinfo {author} {\bibfnamefont {W.}~\bibnamefont {J.}}, \ and\
  \bibinfo {author} {\bibfnamefont {Z.}~\bibnamefont {B.}},\ }\href {\doibase
  doi.org/10.1038/s41566-018-0232-2} {\bibfield  {journal} {\bibinfo  {journal}
  {Nature Photonics}\ }\textbf {\bibinfo {volume} {12}},\ \bibinfo {pages}
  {516} (\bibinfo {year} {2018})}\BibitemShut {NoStop}%
\bibitem [{\citenamefont {Hastings}\ and\ \citenamefont
  {Koma}(2006)}]{Hastings2006}%
  \BibitemOpen
  \bibfield  {author} {\bibinfo {author} {\bibfnamefont {M.~B.}\ \bibnamefont
  {Hastings}}\ and\ \bibinfo {author} {\bibfnamefont {T.}~\bibnamefont
  {Koma}},\ }\href {\doibase 10.1007/s00220-006-0030-4} {\bibfield  {journal}
  {\bibinfo  {journal} {Communications in Mathematical Physics}\ }\textbf
  {\bibinfo {volume} {265}},\ \bibinfo {pages} {781} (\bibinfo {year}
  {2006})}\BibitemShut {NoStop}%
\bibitem [{\citenamefont {Gong}\ \emph {et~al.}(2014)\citenamefont {Gong},
  \citenamefont {Foss-Feig}, \citenamefont {Michalakis},\ and\ \citenamefont
  {Gorshkov}}]{Gong2014}%
  \BibitemOpen
  \bibfield  {author} {\bibinfo {author} {\bibfnamefont {Z.-X.}\ \bibnamefont
  {Gong}}, \bibinfo {author} {\bibfnamefont {M.}~\bibnamefont {Foss-Feig}},
  \bibinfo {author} {\bibfnamefont {S.}~\bibnamefont {Michalakis}}, \ and\
  \bibinfo {author} {\bibfnamefont {A.~V.}\ \bibnamefont {Gorshkov}},\ }\href
  {\doibase 10.1103/PhysRevLett.113.030602} {\bibfield  {journal} {\bibinfo
  {journal} {Phys. Rev. Lett.}\ }\textbf {\bibinfo {volume} {113}},\ \bibinfo
  {pages} {030602} (\bibinfo {year} {2014})}\BibitemShut {NoStop}%
\bibitem [{\citenamefont {Foss-Feig}\ \emph {et~al.}(2015)\citenamefont
  {Foss-Feig}, \citenamefont {Gong}, \citenamefont {Clark},\ and\ \citenamefont
  {Gorshkov}}]{Foss-Feig2015}%
  \BibitemOpen
  \bibfield  {author} {\bibinfo {author} {\bibfnamefont {M.}~\bibnamefont
  {Foss-Feig}}, \bibinfo {author} {\bibfnamefont {Z.-X.}\ \bibnamefont {Gong}},
  \bibinfo {author} {\bibfnamefont {C.~W.}\ \bibnamefont {Clark}}, \ and\
  \bibinfo {author} {\bibfnamefont {A.~V.}\ \bibnamefont {Gorshkov}},\ }\href
  {\doibase 10.1103/PhysRevLett.114.157201} {\bibfield  {journal} {\bibinfo
  {journal} {Phys. Rev. Lett.}\ }\textbf {\bibinfo {volume} {114}},\ \bibinfo
  {pages} {157201} (\bibinfo {year} {2015})}\BibitemShut {NoStop}%
\bibitem [{\citenamefont {Storch}\ \emph {et~al.}(2015)\citenamefont {Storch},
  \citenamefont {van~den Worm},\ and\ \citenamefont {Kastner}}]{Storch_2015}%
  \BibitemOpen
  \bibfield  {author} {\bibinfo {author} {\bibfnamefont {D.-M.}\ \bibnamefont
  {Storch}}, \bibinfo {author} {\bibfnamefont {M.}~\bibnamefont {van~den
  Worm}}, \ and\ \bibinfo {author} {\bibfnamefont {M.}~\bibnamefont
  {Kastner}},\ }\href {\doibase 10.1088/1367-2630/17/6/063021} {\bibfield
  {journal} {\bibinfo  {journal} {New Journal of Physics}\ }\textbf {\bibinfo
  {volume} {17}},\ \bibinfo {pages} {063021} (\bibinfo {year}
  {2015})}\BibitemShut {NoStop}%
\bibitem [{\citenamefont {Matsuta}\ \emph {et~al.}(2017)\citenamefont
  {Matsuta}, \citenamefont {Koma},\ and\ \citenamefont
  {Nakamura}}]{matsuta_improving_2017}%
  \BibitemOpen
  \bibfield  {author} {\bibinfo {author} {\bibfnamefont {T.}~\bibnamefont
  {Matsuta}}, \bibinfo {author} {\bibfnamefont {T.}~\bibnamefont {Koma}}, \
  and\ \bibinfo {author} {\bibfnamefont {S.}~\bibnamefont {Nakamura}},\ }\href
  {\doibase 10.1007/s00023-016-0526-1} {\bibfield  {journal} {\bibinfo
  {journal} {Annales Henri Poincaré}\ }\textbf {\bibinfo {volume} {18}},\
  \bibinfo {pages} {519} (\bibinfo {year} {2017})}\BibitemShut {NoStop}%
\bibitem [{\citenamefont {{Tran}}\ \emph {et~al.}(2018)\citenamefont {{Tran}},
  \citenamefont {{Guo}}, \citenamefont {{Su}}, \citenamefont {{Garrison}},
  \citenamefont {{Eldredge}}, \citenamefont {{Foss-Feig}}, \citenamefont
  {{Childs}},\ and\ \citenamefont {{Gorshkov}}}]{Tran2018}%
  \BibitemOpen
  \bibfield  {author} {\bibinfo {author} {\bibfnamefont {M.~C.}\ \bibnamefont
  {{Tran}}}, \bibinfo {author} {\bibfnamefont {A.~Y.}\ \bibnamefont {{Guo}}},
  \bibinfo {author} {\bibfnamefont {Y.}~\bibnamefont {{Su}}}, \bibinfo {author}
  {\bibfnamefont {J.~R.}\ \bibnamefont {{Garrison}}}, \bibinfo {author}
  {\bibfnamefont {Z.}~\bibnamefont {{Eldredge}}}, \bibinfo {author}
  {\bibfnamefont {M.}~\bibnamefont {{Foss-Feig}}}, \bibinfo {author}
  {\bibfnamefont {A.~M.}\ \bibnamefont {{Childs}}}, \ and\ \bibinfo {author}
  {\bibfnamefont {A.~V.}\ \bibnamefont {{Gorshkov}}},\ }\href@noop {}
  {\bibfield  {journal} {\bibinfo  {journal} {arXiv e-prints}\ ,\ \bibinfo
  {eid} {arXiv:1808.05225}} (\bibinfo {year} {2018})},\ \Eprint
  {http://arxiv.org/abs/1808.05225} {arXiv:1808.05225 [quant-ph]} \BibitemShut
  {NoStop}%
\bibitem [{\citenamefont {Else}\ \emph {et~al.}(2018)\citenamefont {Else},
  \citenamefont {Machado}, \citenamefont {Nayak},\ and\ \citenamefont
  {Yao}}]{else_improved_2018}%
  \BibitemOpen
  \bibfield  {author} {\bibinfo {author} {\bibfnamefont {D.~V.}\ \bibnamefont
  {Else}}, \bibinfo {author} {\bibfnamefont {F.}~\bibnamefont {Machado}},
  \bibinfo {author} {\bibfnamefont {C.}~\bibnamefont {Nayak}}, \ and\ \bibinfo
  {author} {\bibfnamefont {N.~Y.}\ \bibnamefont {Yao}},\ }\href
  {http://arxiv.org/abs/1809.06369} {\bibfield  {journal} {\bibinfo  {journal}
  {arXiv:1809.06369 [cond-mat, physics:physics, physics:quant-ph]}\ } (\bibinfo
  {year} {2018})},\ \bibinfo {note} {arXiv: 1809.06369}\BibitemShut {NoStop}%
\bibitem [{\citenamefont {Verstraete}\ \emph {et~al.}(2008)\citenamefont
  {Verstraete}, \citenamefont {Murg},\ and\ \citenamefont
  {Cirac}}]{Verstraete_2008}%
  \BibitemOpen
  \bibfield  {author} {\bibinfo {author} {\bibfnamefont {F.}~\bibnamefont
  {Verstraete}}, \bibinfo {author} {\bibfnamefont {V.}~\bibnamefont {Murg}}, \
  and\ \bibinfo {author} {\bibfnamefont {J.}~\bibnamefont {Cirac}},\ }\href
  {\doibase 10.1080/14789940801912366} {\bibfield  {journal} {\bibinfo
  {journal} {Advances in Physics}\ }\textbf {\bibinfo {volume} {57}},\ \bibinfo
  {pages} {143} (\bibinfo {year} {2008})},\ \Eprint
  {http://arxiv.org/abs/https://doi.org/10.1080/14789940801912366}
  {https://doi.org/10.1080/14789940801912366} \BibitemShut {NoStop}%
\bibitem [{\citenamefont {Chen}\ \emph {et~al.}(2018)\citenamefont {Chen},
  \citenamefont {Zhou},\ and\ \citenamefont {Xu}}]{chen_measuring_2017}%
  \BibitemOpen
  \bibfield  {author} {\bibinfo {author} {\bibfnamefont {X.}~\bibnamefont
  {Chen}}, \bibinfo {author} {\bibfnamefont {T.}~\bibnamefont {Zhou}}, \ and\
  \bibinfo {author} {\bibfnamefont {C.}~\bibnamefont {Xu}},\ }\href
  {http://stacks.iop.org/1742-5468/2018/i=7/a=073101} {\bibfield  {journal}
  {\bibinfo  {journal} {Journal of Statistical Mechanics: Theory and
  Experiment}\ }\textbf {\bibinfo {volume} {2018}},\ \bibinfo {pages} {073101}
  (\bibinfo {year} {2018})}\BibitemShut {NoStop}%
\bibitem [{\citenamefont {Luitz}\ and\ \citenamefont
  {Bar~Lev}(2019)}]{Luitz2018}%
  \BibitemOpen
  \bibfield  {author} {\bibinfo {author} {\bibfnamefont {D.~J.}\ \bibnamefont
  {Luitz}}\ and\ \bibinfo {author} {\bibfnamefont {Y.}~\bibnamefont
  {Bar~Lev}},\ }\href {\doibase 10.1103/PhysRevA.99.010105} {\bibfield
  {journal} {\bibinfo  {journal} {Phys. Rev. A}\ }\textbf {\bibinfo {volume}
  {99}},\ \bibinfo {pages} {010105} (\bibinfo {year} {2019})}\BibitemShut
  {NoStop}%
\bibitem [{\citenamefont {Lashkari}\ \emph {et~al.}(2013)\citenamefont
  {Lashkari}, \citenamefont {Stanford}, \citenamefont {Hastings}, \citenamefont
  {Osborne},\ and\ \citenamefont {Hayden}}]{lashkari2013}%
  \BibitemOpen
  \bibfield  {author} {\bibinfo {author} {\bibfnamefont {N.}~\bibnamefont
  {Lashkari}}, \bibinfo {author} {\bibfnamefont {D.}~\bibnamefont {Stanford}},
  \bibinfo {author} {\bibfnamefont {M.}~\bibnamefont {Hastings}}, \bibinfo
  {author} {\bibfnamefont {T.}~\bibnamefont {Osborne}}, \ and\ \bibinfo
  {author} {\bibfnamefont {P.}~\bibnamefont {Hayden}},\ }\href {\doibase
  10.1007/JHEP04(2013)022} {\bibfield  {journal} {\bibinfo  {journal} {Journal
  of High Energy Physics}\ }\textbf {\bibinfo {volume} {2013}},\ \bibinfo
  {pages} {22} (\bibinfo {year} {2013})}\BibitemShut {NoStop}%
\bibitem [{\citenamefont {Zhou}\ and\ \citenamefont
  {Chen}(2018)}]{zhou_operator_2018}%
  \BibitemOpen
  \bibfield  {author} {\bibinfo {author} {\bibfnamefont {T.}~\bibnamefont
  {Zhou}}\ and\ \bibinfo {author} {\bibfnamefont {X.}~\bibnamefont {Chen}},\
  }\href {http://arxiv.org/abs/1805.09307} {\bibfield  {journal} {\bibinfo
  {journal} {arXiv:1805.09307 [cond-mat, physics:hep-th]}\ } (\bibinfo {year}
  {2018})},\ \bibinfo {note} {arXiv: 1805.09307}\BibitemShut {NoStop}%
\bibitem [{\citenamefont {Xu}\ and\ \citenamefont
  {Swingle}(2018)}]{xu_locality_2018}%
  \BibitemOpen
  \bibfield  {author} {\bibinfo {author} {\bibfnamefont {S.}~\bibnamefont
  {Xu}}\ and\ \bibinfo {author} {\bibfnamefont {B.}~\bibnamefont {Swingle}},\
  }\href {http://arxiv.org/abs/1805.05376} {\bibfield  {journal} {\bibinfo
  {journal} {arXiv:1805.05376 [cond-mat, physics:hep-th, physics:quant-ph]}\ }
  (\bibinfo {year} {2018})},\ \bibinfo {note} {arXiv: 1805.05376}\BibitemShut
  {NoStop}%
\bibitem [{\citenamefont {Gharibyan}\ \emph {et~al.}(2018)\citenamefont
  {Gharibyan}, \citenamefont {Hanada}, \citenamefont {Shenker},\ and\
  \citenamefont {Tezuka}}]{Gharibyan2018}%
  \BibitemOpen
  \bibfield  {author} {\bibinfo {author} {\bibfnamefont {H.}~\bibnamefont
  {Gharibyan}}, \bibinfo {author} {\bibfnamefont {M.}~\bibnamefont {Hanada}},
  \bibinfo {author} {\bibfnamefont {S.~H.}\ \bibnamefont {Shenker}}, \ and\
  \bibinfo {author} {\bibfnamefont {M.}~\bibnamefont {Tezuka}},\ }\href
  {\doibase 10.1007/JHEP07(2018)124} {\bibfield  {journal} {\bibinfo  {journal}
  {Journal of High Energy Physics}\ }\textbf {\bibinfo {volume} {2018}},\
  \bibinfo {pages} {124} (\bibinfo {year} {2018})}\BibitemShut {NoStop}%
\bibitem [{\citenamefont {Roberts}\ \emph {et~al.}(2018)\citenamefont
  {Roberts}, \citenamefont {Stanford},\ and\ \citenamefont
  {Streicher}}]{roberts2018}%
  \BibitemOpen
  \bibfield  {author} {\bibinfo {author} {\bibfnamefont {D.~A.}\ \bibnamefont
  {Roberts}}, \bibinfo {author} {\bibfnamefont {D.}~\bibnamefont {Stanford}}, \
  and\ \bibinfo {author} {\bibfnamefont {A.}~\bibnamefont {Streicher}},\ }\href
  {\doibase 10.1007/JHEP06(2018)122} {\bibfield  {journal} {\bibinfo  {journal}
  {Journal of High Energy Physics}\ }\textbf {\bibinfo {volume} {2018}},\
  \bibinfo {pages} {122} (\bibinfo {year} {2018})}\BibitemShut {NoStop}%
\bibitem [{\citenamefont {Hohenberg}\ and\ \citenamefont
  {Halperin}(1977)}]{Hohenberg1977}%
  \BibitemOpen
  \bibfield  {author} {\bibinfo {author} {\bibfnamefont {P.~C.}\ \bibnamefont
  {Hohenberg}}\ and\ \bibinfo {author} {\bibfnamefont {B.~I.}\ \bibnamefont
  {Halperin}},\ }\href {\doibase 10.1103/RevModPhys.49.435} {\bibfield
  {journal} {\bibinfo  {journal} {Rev. Mod. Phys.}\ }\textbf {\bibinfo {volume}
  {49}},\ \bibinfo {pages} {435} (\bibinfo {year} {1977})}\BibitemShut
  {NoStop}%
\bibitem [{\citenamefont {Fredrickson}\ and\ \citenamefont
  {Andersen}(1984)}]{Fredrickson1984}%
  \BibitemOpen
  \bibfield  {author} {\bibinfo {author} {\bibfnamefont {G.~H.}\ \bibnamefont
  {Fredrickson}}\ and\ \bibinfo {author} {\bibfnamefont {H.~C.}\ \bibnamefont
  {Andersen}},\ }\href {\doibase 10.1103/PhysRevLett.53.1244} {\bibfield
  {journal} {\bibinfo  {journal} {Phys. Rev. Lett.}\ }\textbf {\bibinfo
  {volume} {53}},\ \bibinfo {pages} {1244} (\bibinfo {year}
  {1984})}\BibitemShut {NoStop}%
\bibitem [{\citenamefont {Chen}\ and\ \citenamefont
  {Zhou}(2018)}]{chen_operator_2018}%
  \BibitemOpen
  \bibfield  {author} {\bibinfo {author} {\bibfnamefont {X.}~\bibnamefont
  {Chen}}\ and\ \bibinfo {author} {\bibfnamefont {T.}~\bibnamefont {Zhou}},\
  }\href {http://arxiv.org/abs/1804.08655} {\bibfield  {journal} {\bibinfo
  {journal} {arXiv:1804.08655 [cond-mat, physics:hep-th, physics:quant-ph]}\ }
  (\bibinfo {year} {2018})},\ \bibinfo {note} {arXiv: 1804.08655}\BibitemShut
  {NoStop}%
\bibitem [{\citenamefont {FISHER}(1937)}]{Fisher1937}%
  \BibitemOpen
  \bibfield  {author} {\bibinfo {author} {\bibfnamefont {R.~A.}\ \bibnamefont
  {FISHER}},\ }\href {\doibase 10.1111/j.1469-1809.1937.tb02153.x} {\bibfield
  {journal} {\bibinfo  {journal} {Annals of Eugenics}\ }\textbf {\bibinfo
  {volume} {7}},\ \bibinfo {pages} {355} (\bibinfo {year} {1937})},\ \Eprint
  {http://arxiv.org/abs/https://onlinelibrary.wiley.com/doi/pdf/10.1111/j.1469-1809.1937.tb02153.x}
  {https://onlinelibrary.wiley.com/doi/pdf/10.1111/j.1469-1809.1937.tb02153.x}
  \BibitemShut {NoStop}%
\bibitem [{\citenamefont {Kolmogorov}\ \emph {et~al.}(1937)\citenamefont
  {Kolmogorov}, \citenamefont {Petrovskii},\ and\ \citenamefont
  {Piskunov}}]{kolmogorov1937study}%
  \BibitemOpen
  \bibfield  {author} {\bibinfo {author} {\bibfnamefont {A.}~\bibnamefont
  {Kolmogorov}}, \bibinfo {author} {\bibfnamefont {I.}~\bibnamefont
  {Petrovskii}}, \ and\ \bibinfo {author} {\bibfnamefont {N.}~\bibnamefont
  {Piskunov}},\ }\href@noop {} {\bibfield  {journal} {\bibinfo  {journal}
  {Selected Works of AN Kolmogorov I}\ ,\ \bibinfo {pages} {248}} (\bibinfo
  {year} {1937})}\BibitemShut {NoStop}%
\bibitem [{\citenamefont {Ablowitz}\ and\ \citenamefont
  {Zeppetella}(1979)}]{ablowitz1979}%
  \BibitemOpen
  \bibfield  {author} {\bibinfo {author} {\bibfnamefont {M.~J.}\ \bibnamefont
  {Ablowitz}}\ and\ \bibinfo {author} {\bibfnamefont {A.}~\bibnamefont
  {Zeppetella}},\ }\href {\doibase 10.1007/BF02462380} {\bibfield  {journal}
  {\bibinfo  {journal} {Bulletin of Mathematical Biology}\ }\textbf {\bibinfo
  {volume} {41}},\ \bibinfo {pages} {835} (\bibinfo {year} {1979})}\BibitemShut
  {NoStop}%
\bibitem [{Note1()}]{Note1}%
  \BibitemOpen
  \bibinfo {note} {At $\alpha =2$, it takes very long time for $\rho (x)$ to
  converge to a Gaussian distribution.}\BibitemShut {Stop}%
\bibitem [{\citenamefont {{Mancinelli, R.}}\ \emph {et~al.}(2002)\citenamefont
  {{Mancinelli, R.}}, \citenamefont {{Vergni, D.}},\ and\ \citenamefont
  {{Vulpiani, A.}}}]{Mancinelli_2002}%
  \BibitemOpen
  \bibfield  {author} {\bibinfo {author} {\bibnamefont {{Mancinelli, R.}}},
  \bibinfo {author} {\bibnamefont {{Vergni, D.}}}, \ and\ \bibinfo {author}
  {\bibnamefont {{Vulpiani, A.}}},\ }\href {\doibase 10.1209/epl/i2002-00251-7}
  {\bibfield  {journal} {\bibinfo  {journal} {Europhys. Lett.}\ }\textbf
  {\bibinfo {volume} {60}},\ \bibinfo {pages} {532} (\bibinfo {year}
  {2002})}\BibitemShut {NoStop}%
\bibitem [{\citenamefont {del Castillo-Negrete}\ \emph
  {et~al.}(2003)\citenamefont {del Castillo-Negrete}, \citenamefont
  {Carreras},\ and\ \citenamefont {Lynch}}]{del-Castillo-Negrete_2003}%
  \BibitemOpen
  \bibfield  {author} {\bibinfo {author} {\bibfnamefont {D.}~\bibnamefont {del
  Castillo-Negrete}}, \bibinfo {author} {\bibfnamefont {B.~A.}\ \bibnamefont
  {Carreras}}, \ and\ \bibinfo {author} {\bibfnamefont {V.~E.}\ \bibnamefont
  {Lynch}},\ }\href {\doibase 10.1103/PhysRevLett.91.018302} {\bibfield
  {journal} {\bibinfo  {journal} {Phys. Rev. Lett.}\ }\textbf {\bibinfo
  {volume} {91}},\ \bibinfo {pages} {018302} (\bibinfo {year}
  {2003})}\BibitemShut {NoStop}%
\bibitem [{\citenamefont {Coulon}\ and\ \citenamefont
  {Roquejoffre}(2012)}]{Coulon_2012}%
  \BibitemOpen
  \bibfield  {author} {\bibinfo {author} {\bibfnamefont {A.-C.}\ \bibnamefont
  {Coulon}}\ and\ \bibinfo {author} {\bibfnamefont {J.-M.}\ \bibnamefont
  {Roquejoffre}},\ }\href {\doibase 10.1080/03605302.2012.718024} {\bibfield
  {journal} {\bibinfo  {journal} {Communications in Partial Differential
  Equations}\ }\textbf {\bibinfo {volume} {37}},\ \bibinfo {pages} {2029}
  (\bibinfo {year} {2012})},\ \Eprint
  {http://arxiv.org/abs/https://doi.org/10.1080/03605302.2012.718024}
  {https://doi.org/10.1080/03605302.2012.718024} \BibitemShut {NoStop}%
\bibitem [{\citenamefont {Brunet}\ and\ \citenamefont
  {Derrida}(1997)}]{Brunet_1997}%
  \BibitemOpen
  \bibfield  {author} {\bibinfo {author} {\bibfnamefont {E.}~\bibnamefont
  {Brunet}}\ and\ \bibinfo {author} {\bibfnamefont {B.}~\bibnamefont
  {Derrida}},\ }\href {\doibase 10.1103/PhysRevE.56.2597} {\bibfield  {journal}
  {\bibinfo  {journal} {Phys. Rev. E}\ }\textbf {\bibinfo {volume} {56}},\
  \bibinfo {pages} {2597} (\bibinfo {year} {1997})}\BibitemShut {NoStop}%
\bibitem [{\citenamefont {Kessler}\ \emph {et~al.}(1998)\citenamefont
  {Kessler}, \citenamefont {Ner},\ and\ \citenamefont {Sander}}]{Kessler_1998}%
  \BibitemOpen
  \bibfield  {author} {\bibinfo {author} {\bibfnamefont {D.~A.}\ \bibnamefont
  {Kessler}}, \bibinfo {author} {\bibfnamefont {Z.}~\bibnamefont {Ner}}, \ and\
  \bibinfo {author} {\bibfnamefont {L.~M.}\ \bibnamefont {Sander}},\ }\href
  {\doibase 10.1103/PhysRevE.58.107} {\bibfield  {journal} {\bibinfo  {journal}
  {Phys. Rev. E}\ }\textbf {\bibinfo {volume} {58}},\ \bibinfo {pages} {107}
  (\bibinfo {year} {1998})}\BibitemShut {NoStop}%
\bibitem [{\citenamefont {Panja}(2004)}]{Panja_2004}%
  \BibitemOpen
  \bibfield  {author} {\bibinfo {author} {\bibfnamefont {D.}~\bibnamefont
  {Panja}},\ }\href {\doibase https://doi.org/10.1016/j.physrep.2003.12.001}
  {\bibfield  {journal} {\bibinfo  {journal} {Physics Reports}\ }\textbf
  {\bibinfo {volume} {393}},\ \bibinfo {pages} {87 } (\bibinfo {year}
  {2004})}\BibitemShut {NoStop}%
\bibitem [{\citenamefont {Brockmann}\ and\ \citenamefont
  {Hufnagel}(2007)}]{Brockmann_2007}%
  \BibitemOpen
  \bibfield  {author} {\bibinfo {author} {\bibfnamefont {D.}~\bibnamefont
  {Brockmann}}\ and\ \bibinfo {author} {\bibfnamefont {L.}~\bibnamefont
  {Hufnagel}},\ }\href {\doibase 10.1103/PhysRevLett.98.178301} {\bibfield
  {journal} {\bibinfo  {journal} {Phys. Rev. Lett.}\ }\textbf {\bibinfo
  {volume} {98}},\ \bibinfo {pages} {178301} (\bibinfo {year}
  {2007})}\BibitemShut {NoStop}%
\end{thebibliography}%

\end{document}